\documentclass[prb,aps,twocolumn,superscriptaddress]{revtex4-1}
\usepackage{graphicx}
\usepackage{dcolumn}
\usepackage{bm}
\usepackage{lipsum}
\usepackage{ragged2e}
\usepackage{braket}
\usepackage{bbm}
\usepackage{color}
\usepackage{xcolor}
\usepackage{amsfonts}
\usepackage{amsmath}
\usepackage{natbib}
\providecommand{\keywords}[1]{\textbf{\textit{Index terms---}} #1}
\usepackage[mathlines]{lineno}
\usepackage[colorinlistoftodos]{todonotes}
\usepackage[colorlinks=true, allcolors=blue]{hyperref}

\begin{document}

\newcommand{\unamone}{Departamento de Sistemas Complejos, Instituto de Fisica,
Universidad Nacional Aut\'onoma de M\'exico, Apartado Postal 20-364,01000,
Ciudad de M\'exico, M\'exico.}
\newcommand{\unamtwo}{Instituto de F\'isica,
Universidad Nacional Aut\'onoma de M\'exico, Apartado Postal 20-364 01000,
Ciudad de  M\'{e}xico, M\'exico}
\newcommand{\uabc}{Facultad de Ciencias, Universidad Aut\'onoma de Baja California, Apartado Postal 1880, 22800 Ensenada, Baja California, M\'exico}

\title{Dynamical Floquet spectrum of Kekul\'e-distorted graphene under normal incidence of electromagnetic radiation}

\author{M. A. Mojarro}
\affiliation{\uabc}
\author{V. G. Ibarra-Sierra}
\email{vickkun@fisica.unam.mx}
\affiliation{\unamone}
\author{J. C. Sandoval-Santana}
\affiliation{\unamtwo}
\author{R. Carrillo-Bastos}
\affiliation{\uabc}
\author{Gerardo G. Naumis}
\affiliation{\unamone}

\date{\today}

\begin{abstract}

Electromagnetic dressing by a high-frequency field drastically modifies the electronic transport properties on Dirac systems. Here its effects on the energy spectrum of graphene with two possible phases of  Kekul\'e distortion (namely, Kek-Y and Kek-O textures) are studied. Using Floquet theory it is shown how circularly polarized light modifies the gapless spectrum of the Kek-Y texture, producing dynamical band gaps at the Dirac point that depends on the amplitude and the frequency of the electric field, and breaks the valley degeneracy of the gapped spectrum of the Kek-O texture. To further explore the electronic properties under circularly polarized radiation,  the dc conductivity is studied by using the Boltzmann approach and considering both inter-valley and intra-valley contributions.
When linearly polarized light is considered, the band structure of both textures is always modified in a perpendicular direction to the electric field. While the band structure for the Kek-Y texture remains gapless, the gap for the Kek-O texture is reduced considerably. For this linear polarization it is also shown that non-dispersive bands can appear by a precise tuning of the light field parameters thus inducing dynamical localization. The present results suggest that optical measurements will allow to distinguish between different Kekul\'e bond textures.

\end{abstract}

\keywords{Suggested keywords}

\maketitle

\section{Introduction}
Due to its hexagonal symmetry, graphene possess a double cone linear spectrum\cite{graphene}, each of them labeled by $K$ and $K'$ at the corners of the corresponding hexagonal Brillouin zone. As they are separated by a large momentum, these two nonequivalent cones can be considered independent and be described by an spin-like degree of freedom: the valley isospin; this provided that any perturbation in the system is larger when compared with the lattice parameter\cite{katsnelson2012graphene}. There are several mechanisms that allow to engineer the spectrum of graphene, these include interactions with substrates\cite{zhou2007substrate}, strain\cite{vozmediano2010gauge,amorim2016novel,naumis2017electronic}, moire patterns\cite{ponomarenko2013cloning}, adatoms\cite{Bianchi2010, Kaasbjerg2019}, magnetic fields\cite{novoselov2004electric,jiang2007infrared,guinea2006electronic}, and time dependent electromagnetic fields\cite{eckardt2015high,LaserGap,FloquetTop}. In this manuscript, we study the effect on the band spectrum of the combinations of two of these mechanisms. Inspired by the recent experimental confirmation of a Kekul\'{e} Y-shaped (Kek-Y) phase in graphene when deposited on a Cooper substrate\cite{gutierrez}, and the results of density functional theory calculations that suggest the possibility of obtaining the Kekul\'{e} O-shaped  (Kek-O) phase by depositing graphene on top of a topological insulator\cite{lin2017competing,tajkov2019uniaxial}, we explore the effect of irradiating Kekul\'e distorted graphene with polarized light (linearly and circularly) at normal incidence. Also, we calculate the dc conductivity using a Boltzmann formalism\cite{Mahan1990,Rossiter1991}, which could be suitable to compare with experiments.

In graphene, we call a Kekul\'{e} distortion to a periodic bond distance modification (local strain) with a spatial frequency that increases the size of the unit cell to that of an hexagonal ring of carbon atoms\cite{hou2007electron}. This results in the merging of the two Dirac cones at the center of the Brillouin zone, producing either a gap (Kek-O) or the superposition of two cones with different Fermi velocities (Kek-Y)\cite{gamayun}. There has been several works exploring the consequences of a Kekul\'{e} texture on graphene, specially after the experimental realization of Gutierrez, \textit{et. al}\cite{gutierrez}: Gamayun \textit{et. al} demostrated the absence of a gap for a Kek-Y distortion and deduced the Hamiltonian for both types of distortions\cite{gamayun}. Andrade \textit{et. al} studied the effects of uniaxial strain\cite{strainKek}, which previously was shown to affect the formation of the Kekul\'{e} pattern\cite{formation-strain-kek}. Other works have investigated the electronic transport properties of Kekul\'{e} distorted graphene\cite{KuboKek,GNRKek,Barrier-Kek,kek-device}, the competition with spin-orbit interactions\cite{kek-soi}, as well as the consequences of a Kekul\'{e} distortion in analogue systems, where low energy excitations are phonons\cite{phononsKek} or magnons\cite{magnonKek,moulsdale2019unconventional}. 

It is well established that electromagnetic radiation can dramatically change the band structure of an electronic system\cite{AnnualReview-Oka,rudner2019floquet,Kibis2014,Morina2DEG}. In particular, for Dirac-like systems (with linear dispersion), it may lead to the creation of gaps and changes in their topological flavor.
One of the first studies addressing the manipulation of electronic transport by electromagnetic fields in the so-called Dirac Matter is the one by Fistul and Efetov \cite{Fistul-PRL98}, they suggested that the \textit{dynamic gap} induced by irradiating graphene with an electromagnetic field can serve as a way to control the Klein tunneling in a pn-junction\cite{Efetov-prb2008}. Later, T. Oka and H. Aoki pointed out the non-trivial character of this gap, and therefore predicted a photoinduced dc Hall current associated with it\cite{Takashi-2009}. The presence of this gap can be demonstrated analytically\cite{grapheneIrradiated1,grapheneIrradiated2,kibis} and be confirmed using the standard quantum-field theory approach, where electron-photon interaction in graphene irradiated by polarized photons results in a metal-insulator transition\cite{Kibis2}. These results extends to other Dirac materials\cite{polarizability,NonLinearEff}, like topological insulators\cite{Yudin2016}, borophene\cite{boropheneIrradiated1,boropheneIrradiated2,Sandoval2020FloquetSpectrum,kunold2020floquet}, $\alpha-\mathcal{T}_3$ graphene \cite{alphaIrradiated1,alphaIrradiated2,alphaIrradiated3,alphaTransitions} and silicene \cite{Ezawa-2013}. Moreover, although the electronic and optical conductivity that results from the application of an in-plane electromagnetic field has already been analytically studied\cite{KuboKek}, until now the dynamical band structure of irradiated Kekul\'{e} graphene, where the two valley are nested in the same point, has not been explored yet.

In this paper, we address the general problem of an electron in a Kekul\'e distorted graphene under circularly and linearly polarized light. The circularly polarized light problem is addressed in the weak field regime, and the corresponding linear light problem is solved in the high-frequency regime. We report the quasienergy spectrum for both textures, Kek-Y and Kek-O, considering these two types of polarization. In the case of circularly polarized light, we show the exact expressions for gap opening conditions. For linearly polarization light, we demonstrate that it breaks the angular symmetry in the quasienergy spectrum. To understand the physical properties of this system, we calculate the dc conductivity by using the Boltzmann approach and considering inter-valley and intra-valley contributions under circularly polarized light.

The paper is organized as follows. In Sec. \ref{sec:Sec2} we introduce the honeycomb lattice of Kekul\'e distorted graphene, as well as its low-energy Hamiltonian, and in Sec. \ref{sec:Sec3} we compute the quasienergy spectrum by solving the Dirac equation when an electromagnetic wave is applied normally to the lattice. Finally, in Sec. \ref{sec:Sec4} we calculate the dc conductivity of Kek-Y distorted graphene under a circularly polarized electromagnetic wave, and in Sec. \ref{sec:Sec5} we present the conclusions of this work.

\begin{figure}[t]
\begin{center}\quad\qquad\large(a)\qquad\qquad\qquad\qquad\qquad\large(b)\end{center}
\includegraphics[width=.295\textwidth]{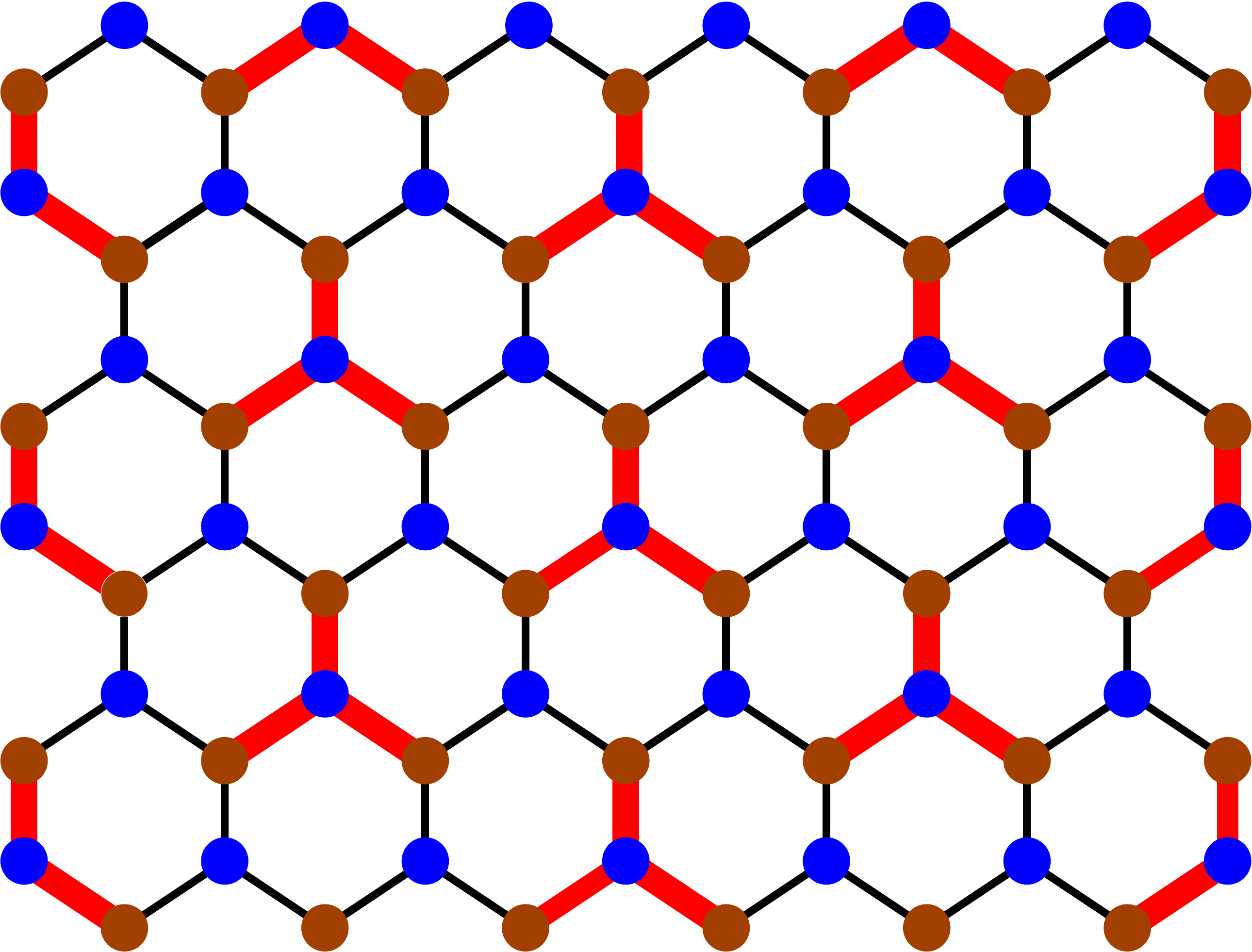}
\includegraphics[width=.168\textwidth]{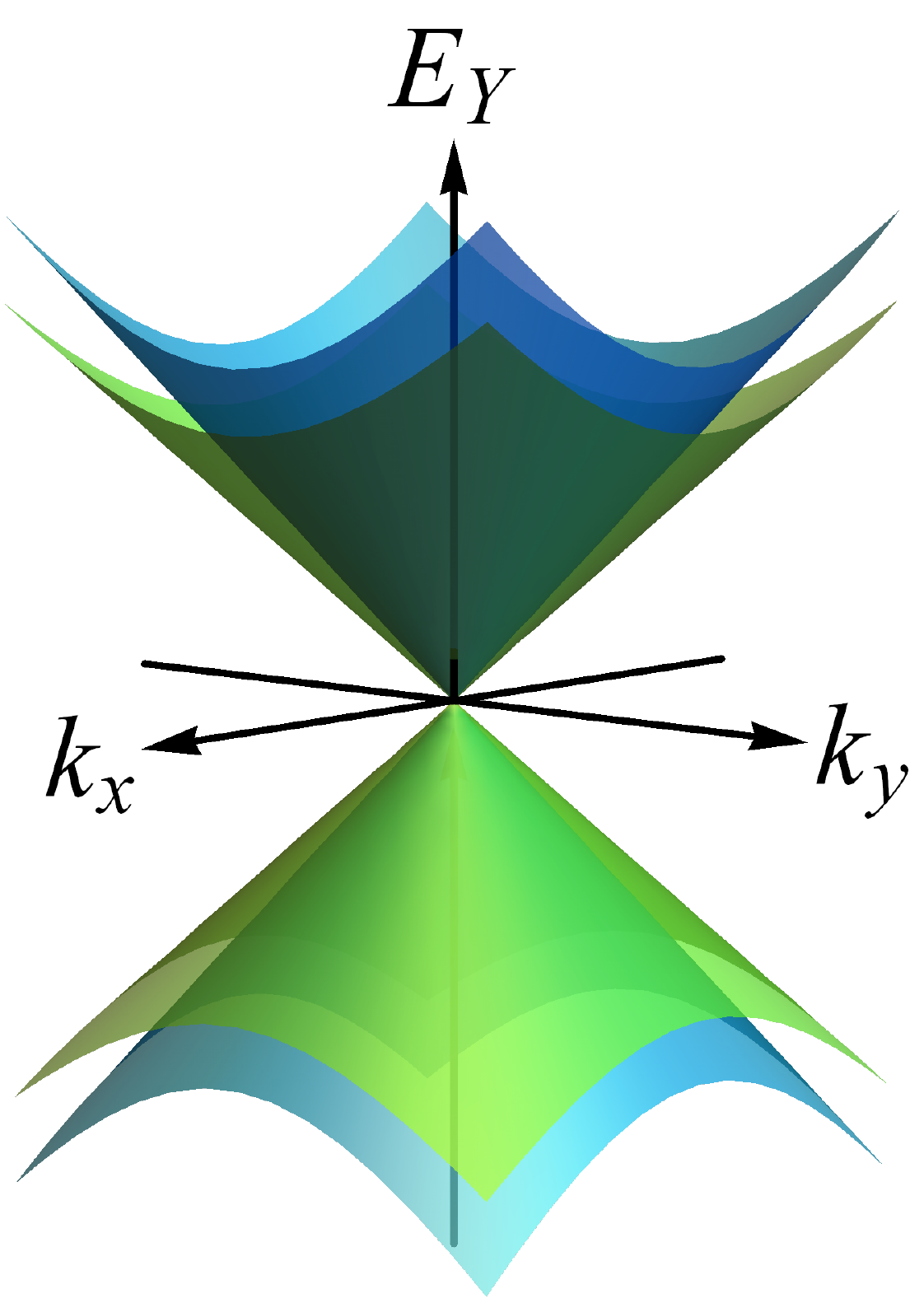}\\
\begin{center}\quad\qquad\large(c)\qquad\qquad\qquad\qquad\qquad\large(d)\end{center}
\vspace{-0.5cm}
\includegraphics[width=.295\textwidth]{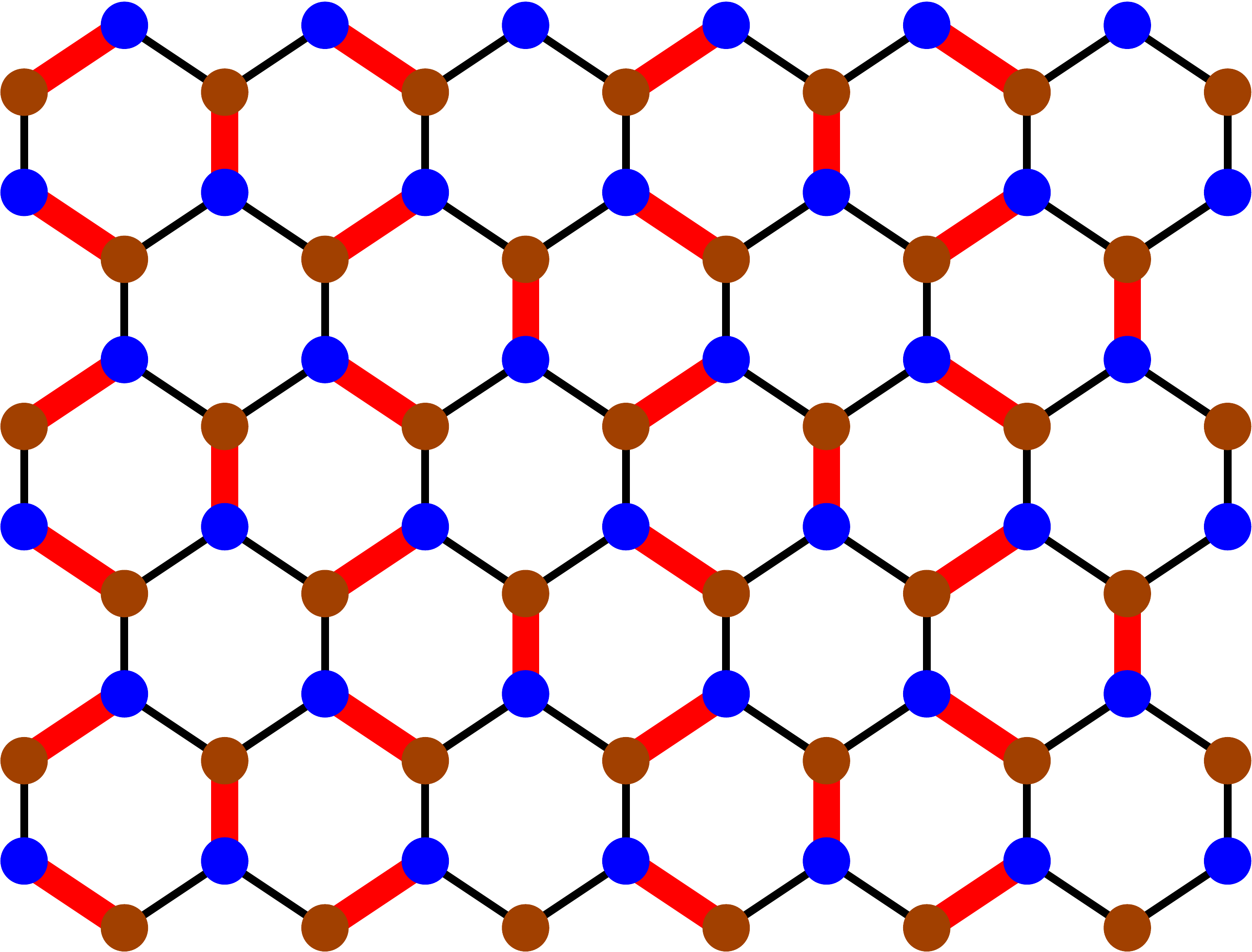}
\includegraphics[width=.138\textwidth]{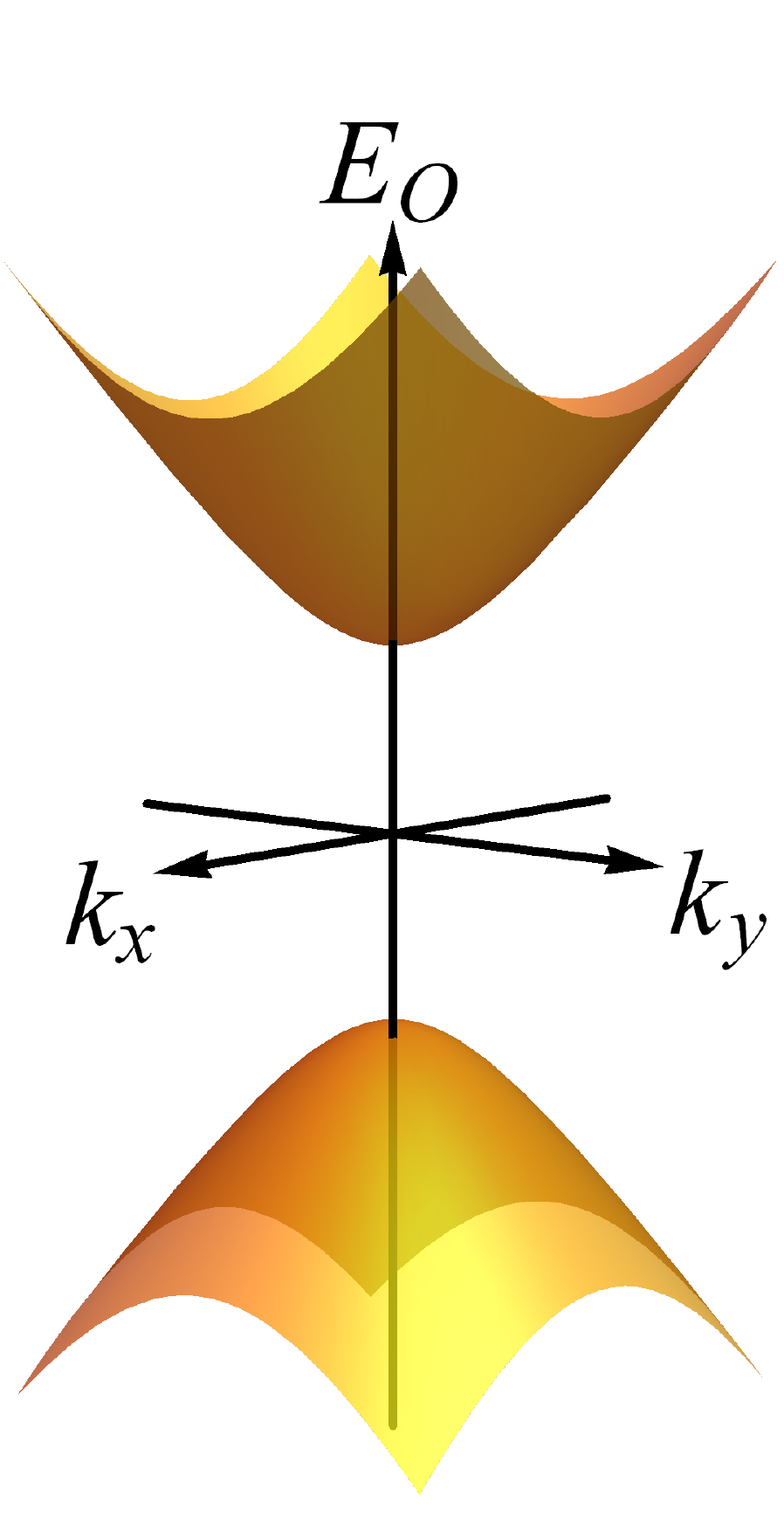}
\caption{\label{fig:KekO_KekY} Kekul\'e distorted honeycomb lattice and dispersion relation for (a) the Kek-Y ($\nu=1$) and (c) the Kek-O ($\nu=0$) texture. The red and black lines represent slightly shorter and longer bond lengths, respectively. For $\nu=1$, the internal and external cones touch each other at the Dirac point, while for $\nu=0$ a gap is open, as shown in (b) and (d), respectively.}
\end{figure}

\section{The continuum Hamiltonian for Kekul\'{e}-distorted graphene}\label{sec:Sec2}
Considering a monolayer of Kekul\'e-distorted graphene lying in the plane  at $z=0$, we define the lattice vectors: $\boldsymbol{a}_1=\boldsymbol{\delta}_3-\boldsymbol{\delta}_1$ and $\boldsymbol{a}_2=\boldsymbol{\delta}_3-\boldsymbol{\delta}_2$, in terms of the three nearest neighbors vectors:  $\boldsymbol{\delta}_1=\frac{1}{2}(\sqrt{3},-1)a_0$, $\boldsymbol{\delta}_2=-\frac{1}{2}(\sqrt{3},1)a_0$, $\boldsymbol{\delta}_3=(0,1)a_0$ (with the bond length $a_0\approx$1.42\AA). Thus Kekul\'e distortions can be described, in the first neighbor tight-binding Hamiltonian\cite{gamayun}, 
\begin{equation}\label{TightBinding}
    H_{tb}=- \sum_{\boldsymbol{r},l}t_{\boldsymbol{r},l}a_{\boldsymbol{r}}^{\dagger}b_{\boldsymbol{r}+\boldsymbol{\delta}_l}+\mathrm{H}.\mathrm{c}.,
\end{equation}
by  a bond-density wave
\begin{equation}
    t_{\boldsymbol{r},l}/t_0=1+2\:\Re\Big\{\tilde{\Delta}e^{i(p\textbf{K}_++q\textbf{K}_-)\cdot\boldsymbol{\delta}_l+i\textbf{G}\cdot\boldsymbol{r}-i2\pi (p+q)/3}\Big\},
\end{equation} 
which modifies periodically the hopping amplitudes, $t_{\boldsymbol{r},l}$, between an atom at site $\boldsymbol{r}=n_1\boldsymbol{a}_1+n_2\boldsymbol{a}_2$ ($n_1, n_2 \in \mathbb{Z}$) and its three nearest-neighbor sites at $\boldsymbol{r}+\boldsymbol{\delta}_l$. There, $t_0\approx2.7$ eV is the hopping amplitude of the unperturbed C-C bond, $\tilde{\Delta}=e^{i2\pi (p+q+m)/3}\Delta_0$ ($p, q \in\mathbb{Z}_{3}$ and $m\in\mathbb{Z}$) is the so-called Kekul\'e parameter\cite{gamayun},  and $\textbf{G}=\textbf{K}_+-\textbf{K}_-$ is the Kekul\'e wave vector, with $\textbf{K}_\pm=\frac{2\pi}{9}\sqrt{3}(\pm1,\sqrt{3})a_0$ the reciprocal lattice vectors. We can distinguish between the Kek-O and the Kek-Y textures through the index $\nu=1+q-p$ mod $3$, where $|\nu|=1$ accounts for Kek-Y and $\nu=0$ for Kek-O (see Fig. \ref{fig:KekO_KekY}).

Using the low-energy approximation, it is possible to show that the corresponding continuum Hamiltonian for Kekul\'e-distorted graphene is \cite{gamayun} 
\begin{equation}\label{LowEnergy}
H_0=
\begin{pmatrix}
v \boldsymbol{\sigma}\cdot\boldsymbol{p} & \tilde{\Delta} Q_{\nu} \\
\tilde{\Delta}^\ast Q_{\nu}^\dagger & v\boldsymbol{\sigma}\cdot\boldsymbol{p}
\end{pmatrix},
\end{equation}
where  $\boldsymbol{p}=\hbar(k_x,k_y)$ is the momentum, $\boldsymbol{\sigma}=(\sigma_x,\sigma_y)$ is the Pauli vector, with the Pauli matrices $\sigma_i$ acting on the pseudospin degree of freedom; while the matrix is expand on the valley degree of freedom,  such that the valley mixing operator $Q_\nu$ is defined by
\begin{equation}\label{Qoperator} 
Q_\nu=\left\{\begin{aligned}&3t_0\sigma_z \qquad\qquad\qquad \mathrm{if}\quad \nu=0,\\ &v_F(\nu p_x-ip_y)\sigma_0 \quad \mathrm{if}\quad |\nu|=1,\end{aligned} \right. \end{equation}
with $v=\frac{3}{2}t_0a_0/\hbar\approx c/300$ the Fermi velocity in pristine graphene. 

The electronic band structure is obtained by solving the eigenvalue problem $H\boldsymbol{\Psi}=E\boldsymbol{\Psi}$ in  momentum space, where $\boldsymbol{\Psi}$ is a four-component spinor, which contains the amplitudes on  sublattices $A$ and $B$ for the valleys $K$ and $K'$. Therefore the spectrum for graphene with Kek-Y distortion consists of two concentric gapless Dirac cones  (see Fig. \ref{fig:KekO_KekY}(b)),
\begin{equation}\label{kekY}
E_{\mathrm{Y}}^{\xi,\eta}(k)=\eta v\hbar k(1+\xi\Delta_0),
\end{equation}
where $k=\sqrt{k_x^2+k_y^2}$ is the total momentum, and the corresponding spinors, that depend only on the momentum direction, $\theta=\tan^{-1}(k_y/k_x)$, are given by
\begin{equation}\label{baseSatateSpinors}
    \boldsymbol{\Psi}_{\xi,\eta}(\boldsymbol{k})=\frac{1}{2}\big(\xi e^{-i2\theta},\:\xi\eta e^{-i\theta},\:\eta e^{-i\theta},\:1\big)^{\mathrm{T}}.
\end{equation}
For graphene with Kek-O distortion, two degenerate gapped cones are found (see Fig. \ref{fig:KekO_KekY}(d)),
\begin{equation}
    E^{\eta}_{\mathrm{O}}(k)=\eta\sqrt{(v\hbar k)^2+(3t_0\Delta_0)^2}.
\end{equation} 
Here $\eta=\pm$ denotes the band (conduction or valence, respectively) and $\xi=\pm$ correspond to the cone (internal or external, respectively).
The two concentric Dirac cones in Kek-Y spectrum are characterized by two different velocities, $v(1+\Delta_0)$ for the internal cone, and $v(1-\Delta_0)$ for the external cone. 

\section{Kekul\'e-distorted graphene under electromagnetic radiation}\label{sec:Sec3}

To study the dynamics of charge carriers in Kekul\'e-distorted graphene under electromagnetic radiation, we introduce a minimal coupling $\boldsymbol{p}\rightarrow\boldsymbol{\pi}=\boldsymbol{p}-e\boldsymbol{A}$ in the low-energy Hamiltonian \eqref{LowEnergy}, where $\boldsymbol{A}=(A_x,A_y)$ is the vector potential of the electromagnetic wave, which is a periodic function of time, and $e$ the electron charge. Therefore, from the Eq. \eqref{LowEnergy} we obtain
\begin{equation}\label{GeneralHam}
H(t)=
\begin{pmatrix}
v \boldsymbol{\sigma}\cdot\boldsymbol{\pi} & \tilde{\Delta}Q_{\nu}(t) \\
\tilde{\Delta}^\ast Q_{\nu}^\dagger(t) & v\boldsymbol{\sigma}\cdot\boldsymbol{\pi}
\end{pmatrix},
\end{equation}
with $Q_{\nu}(t)=v_F(\nu \pi_x-i\pi_y)\sigma_0$ for $|\nu|=1$,
where $\pi_x=p_x-eA_x$ and $\pi_y=p_y-eA_y$. For $\nu=0$ the operator $Q_\nu$ remains invariant (see Eq. \eqref{Qoperator}). The Dirac equation for charge carries is thus given by  
\begin{equation}\label{DiracEquation}
i\hbar\frac{d}{dt}\boldsymbol{\Psi}(\boldsymbol{k},t)=H(t)\boldsymbol{\Psi}(\boldsymbol{k},t),
\end{equation}
where $\boldsymbol{\Psi}(\boldsymbol{k},t)$ is a four-component spinor. We can write the Hamiltonian \eqref{GeneralHam} as follow
\begin{equation}
    H(t)=H_0+V(t),
\end{equation}
where $H_0$ is given in Eq. \eqref{LowEnergy}, and 
\begin{equation}\label{perturbation}
    V(t)=\begin{pmatrix}
    -ev\boldsymbol{\sigma}\cdot\boldsymbol{A}&\tilde{\Delta}W_\nu\\
    \tilde{\Delta}^{\ast}W^{\dagger}_\nu&-ev\boldsymbol{\sigma}\cdot\boldsymbol{A}
    \end{pmatrix},
\end{equation}
is the external perturbation due to the presence of the electromagnetic wave, where $W_\nu=-ev(\nu A_x-iA_y)\sigma_0$ if $|\nu|=1$, and $W_\nu=0$ if $\nu=0$.

We are interested in deducing the analytical expression of the Floquet dynamical spectrum of charge carriers. Therefore, instead of following the standard perturbation theory, we make the following ansatz \cite{kibis,Kibis2,Kibis2017AllOptical}
\begin{equation}\label{ansatz}
    \boldsymbol{\Psi}(\boldsymbol{k},t)=e^{-i\varepsilon t/\hbar}\sum_{n=1}^4a_{n}(\boldsymbol{k},t)\boldsymbol{\psi}_n(t),
\end{equation}
where the quasienergy $\varepsilon$,  and the time-dependent coefficients $a_{n}(\boldsymbol{k},t)$ are to be determined; while 
the four-component spinor $\boldsymbol{\psi}_n(t)$ ($n\in\{1,2,3,4\}$) is the $n$-th solution of the matrix differential equation,
\begin{equation}\label{NonStatDiracE}
i\hbar\frac{d}{dt}\boldsymbol{\psi}_n(t)=V(t)\boldsymbol{\psi}_n(t),
\end{equation}
with $V(t)$ defined by Eq. \eqref{perturbation}. In the ansatz of Eq. \eqref{ansatz}, dependence on momentum is contained in the coefficients  $a_{n}(\boldsymbol{k},t)$ and, according to the Floquet theory\cite{Shirley1965Solution,sandoval2019method}, $\varepsilon$ describes the dynamical spectrum  of quantum systems exposed to periodic perturbations in time. In the subsequent sections we obtain analytical expressions for the quasienergies $\varepsilon$ considering normal incidence of electromagnetic radiation with both circular and linear polarization.

\subsection{Circularly polarized light}

Consider the normal incidence of a circularly polarized electromagnetic wave defined by the vector potential, 
\begin{equation}\label{potential}
    \boldsymbol{A}=\frac{E_0}{\Omega}\big(\cos{(\Omega t)},\:\sin{(\Omega t)}\big),
\end{equation}
where $E_0$ is the amplitude of the electric field, taken as constant, and $\Omega$ is the angular frequency. Also, we have neglected the third dimension. The corresponding electric field is given by $\boldsymbol{E}=-\partial\boldsymbol{A}/\partial t=E_0\big(\sin{(\Omega t)},\:-\cos{(\Omega t)}\big)$. In the following subsections, we analyze the resulting spectrum for the two kinds of Kekul\'e bond textures.

\subsubsection{Kek-Y texture under circularly polarized radiation}
For simplicity, in the Hamiltonian for Kek-Y distorted graphene we take $\nu=1$ and a real $\tilde{\Delta}=\Delta_0$; the case with $\nu=-1$ and a complex $\tilde{\Delta}$ can be obtained by an unitary transformation \cite{gamayun}. 

Here it is convenient to define 
\begin{equation}\label{aproxx}
    \tilde{E}=\frac{eE_0v}{\hbar\Omega^2}. 
\end{equation}
Since for this case we are interested in the effects of a weak electromagnetic field, we ask $ \tilde{E}\ll1$ such that,
\begin{equation}
    \left(\frac{ev}{\Omega}\right)E_0\ll\hbar\Omega,
\end{equation}
which means that the interaction energy between the electric field and the induced dipole moment $ev/\Omega$, is smaller than the energy of a photon.

\begin{figure}[t]
\begin{flushleft}\qquad\quad\qquad\large(a)\quad\qquad\qquad\qquad\qquad\large(b) \end{flushleft}
\includegraphics[width=.205\textwidth]{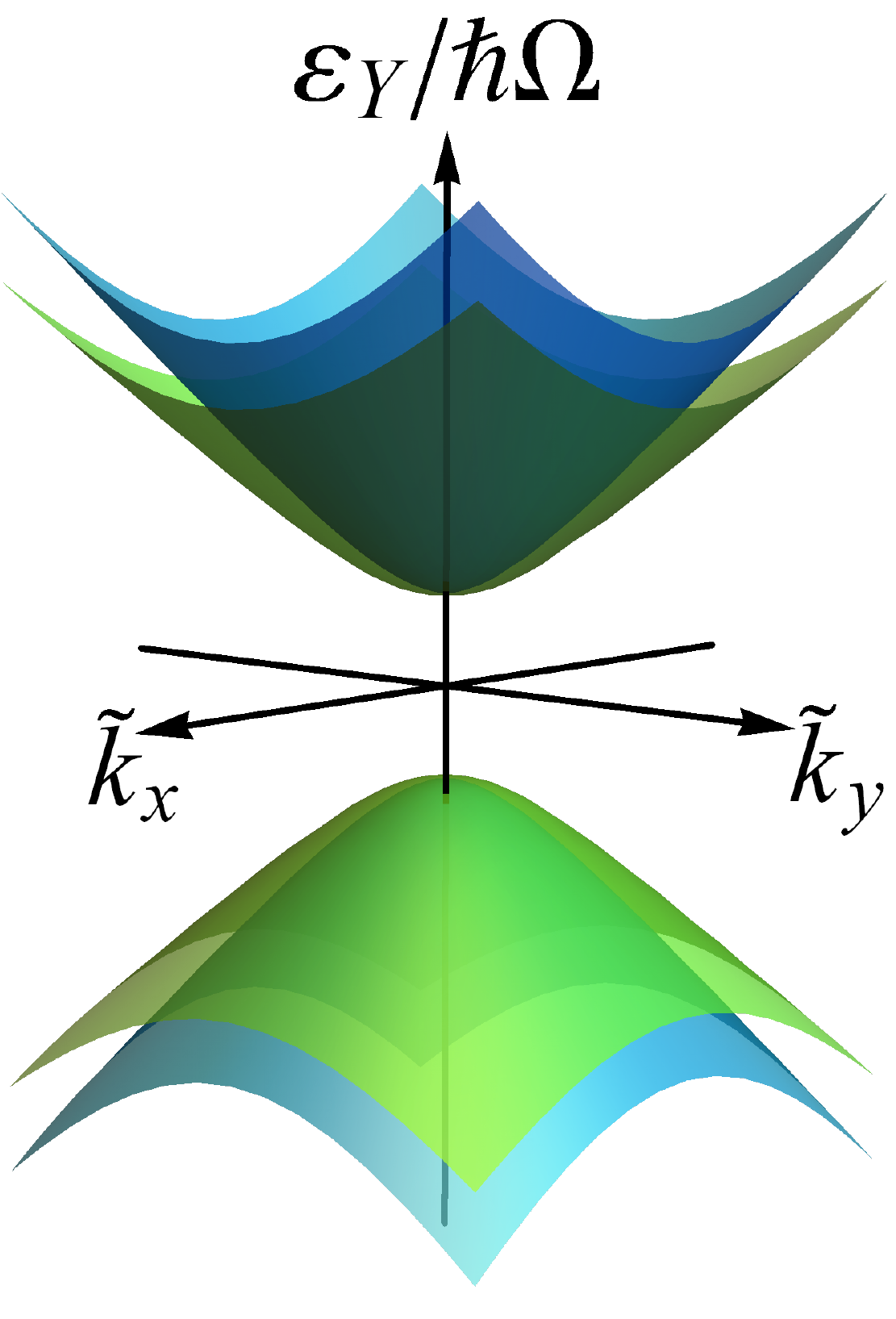}
\includegraphics[width=.272\textwidth]{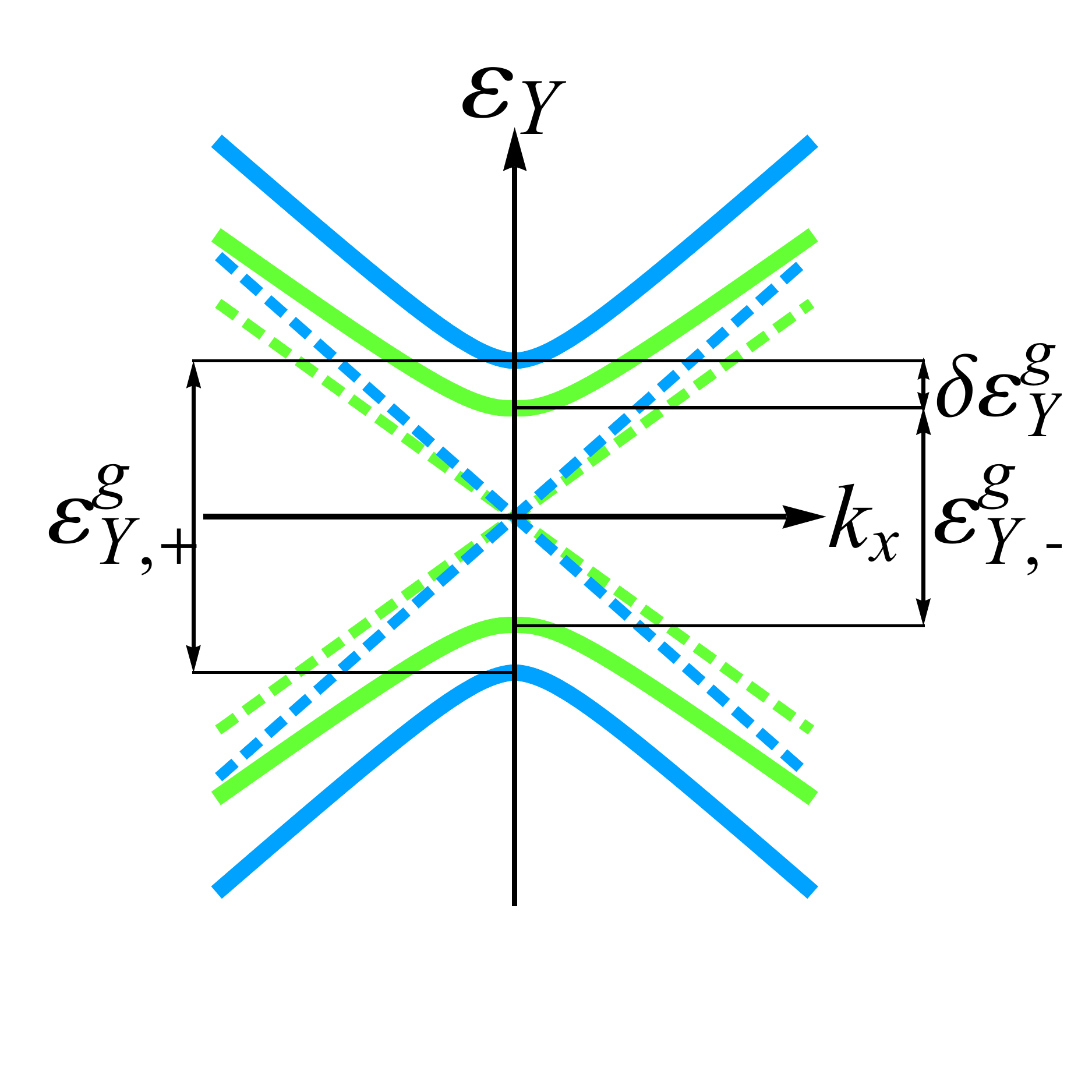}\\
\begin{flushleft}\qquad\quad\qquad\large(c)\qquad\qquad\quad\qquad\qquad\large(d) \end{flushleft}
\includegraphics[width=.205\textwidth]{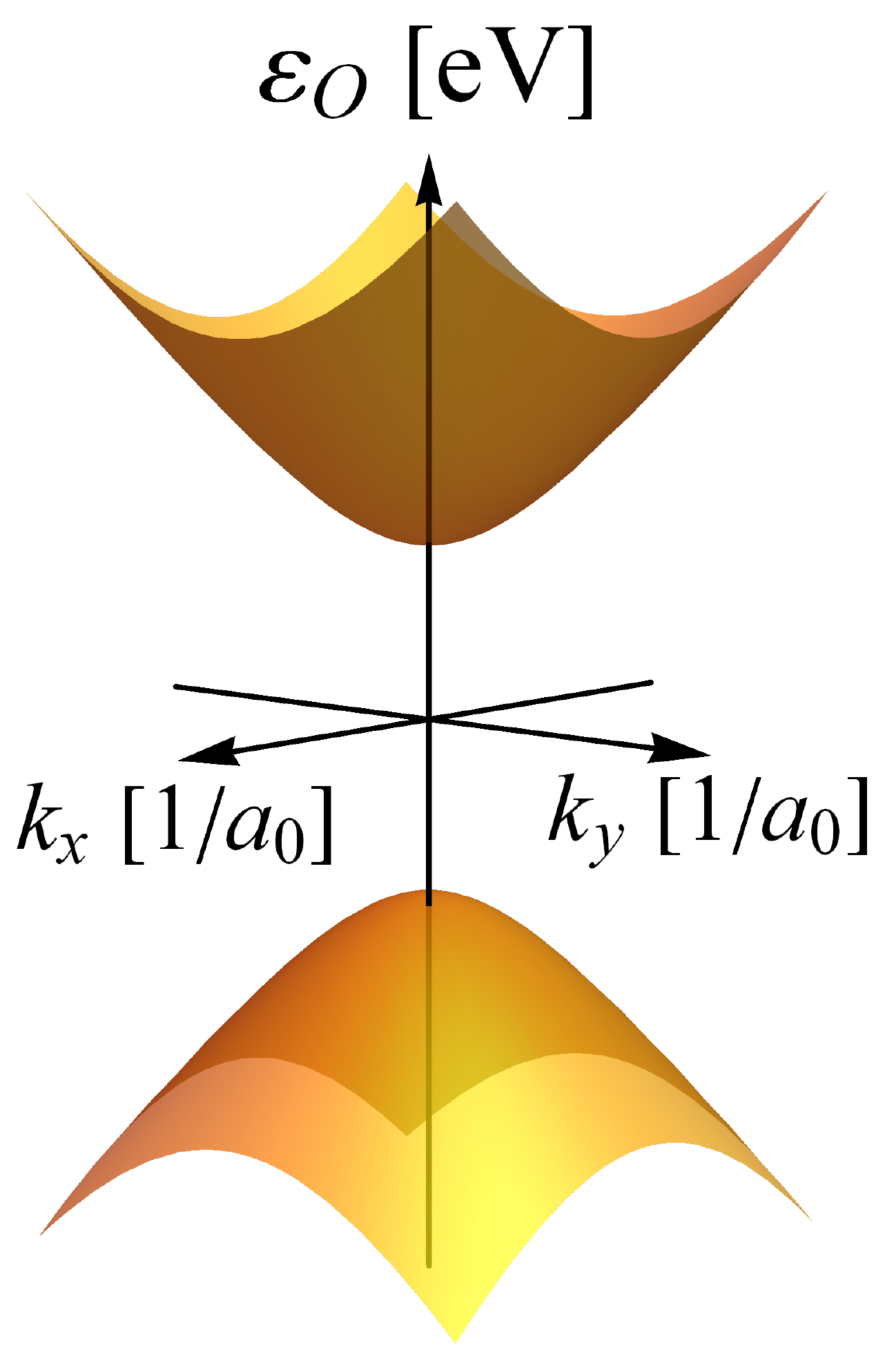}
\includegraphics[width=.272\textwidth]{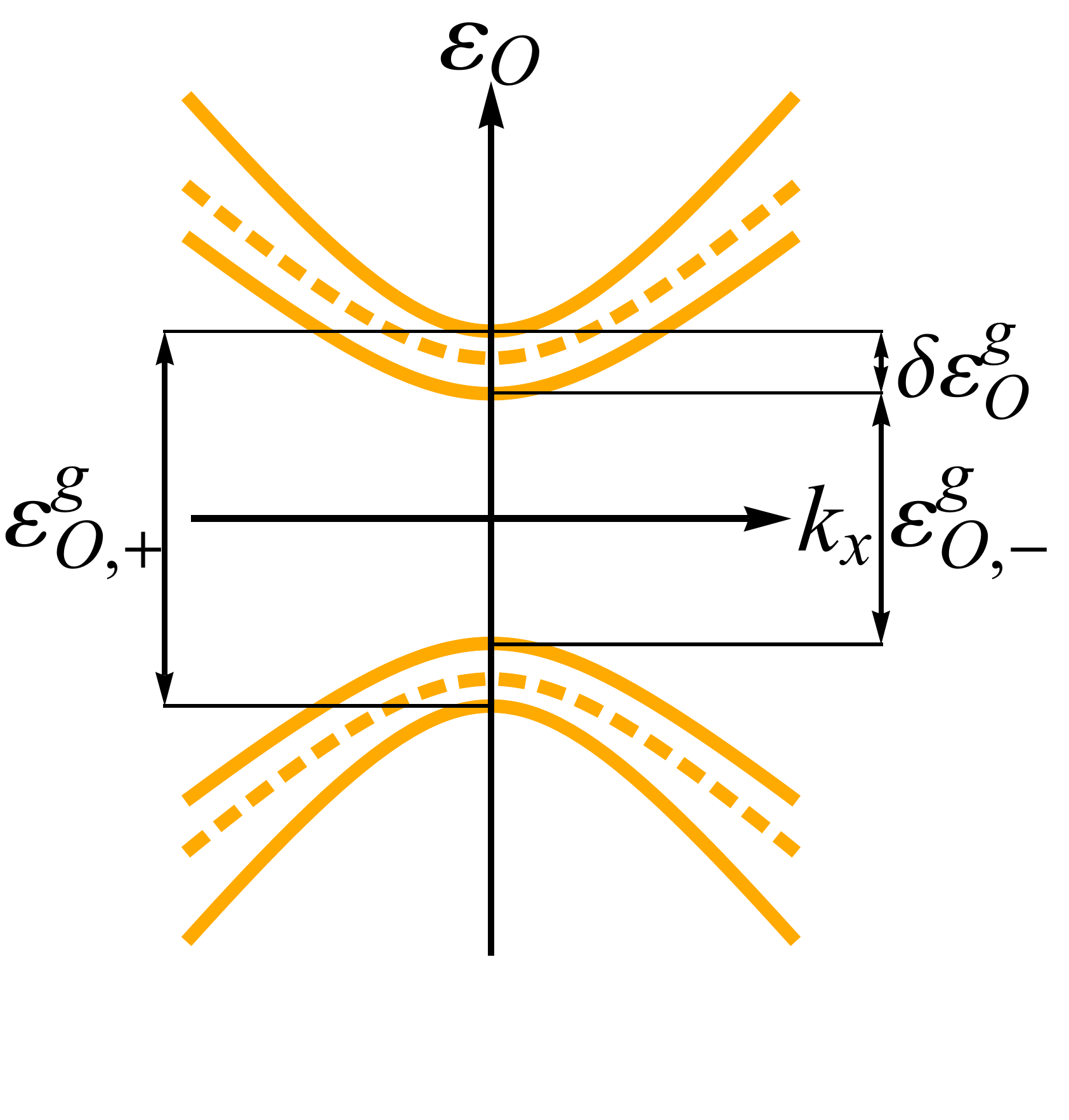}
\caption{\label{fig:KekCirc} Quasienergy spectrum for (a) Kek-Y ($\nu=1$) and (c) Kek-O ($\nu=0$) textured graphene, irradiated with a weak circularly polarized electromagnetic wave, such that $\tilde{E}\ll1$. Here we have defined $\tilde{k}_x\equiv k_xv/\Omega$ and $\tilde{k}_y\equiv k_yv/\Omega$. As shown schematically in (b) and (d) for a cut along the direction $k_y=0$, two quasienergy band gaps appear at the Dirac point for the Kek-Y texture, and the valley degeneracy breaks for the Kek-O texture, respectively. The dotted lines represent the energy bands in absence of the external electromagnetic wave.  
Considering an amplitude of the electric field $E_0=9$ V/m, and a frequency $\Omega=0.5$ THz, the quasienergy band gaps are  $\varepsilon_{\mathrm{Y},-}^{g}\approx1.9442$ $\mu$eV and $\varepsilon_{\mathrm{Y},+}^{g}\approx1.9835$ $\mu$eV, for the Kek-Y texture, and $\varepsilon_{\mathrm{O},-}^{g}\approx1.6201341$ eV and $\varepsilon_{\mathrm{O},+}^{g}\approx1.620138$ eV for the Kek-O texture, with $\delta\varepsilon_{\mathrm{Y}}^{g}\approx19.697$ $n$eV and $\delta\varepsilon_{\mathrm{O}}^{g}\approx1.9639$ $\mu$eV.}
\end{figure}

Under these considerations, we obtain analytical solutions of Eq. \eqref{DiracEquation}. To this end, we first find the vector solutions $\boldsymbol{\psi}_n$ of  Eq. \eqref{NonStatDiracE}; later we build  the ansatz of Eq. \eqref{ansatz} and substitute it into Eq. \eqref{DiracEquation}, to finally find the coefficients 
$a_{n}(\boldsymbol{k},t)$ and the energy eigenvalues $\varepsilon$ (see Appendix \ref{Append-A}). Therefore, the quasienergy spectrum for Kek-Y distorted graphene irradiated with circularly polarized light is
\begin{eqnarray}\label{QuasiYCir}
   \nonumber \varepsilon_{\mathrm{Y}}^{\xi,\eta}(k)=\eta\frac{1}{2}\Big(&&\sqrt{(\alpha-\hbar\Omega)^2+(2v\hbar k)^2}\\
    &&+\:\xi\sqrt{(\beta-\hbar\Omega)^2+(2\Delta_0 v\hbar k)^2}\Big),
\end{eqnarray}
where 
\begin{eqnarray}
     \alpha&=&\hbar\Omega\sqrt{1+(2\tilde{E})^2},\\
     \beta&=&\hbar\Omega\sqrt{1+(2\Delta_0\tilde{E})^2}.
\end{eqnarray}
In Appendix \ref{Append-A}, we show the corresponding four-component spinors. The resulting quasienergy spectrum is shown in Fig. \ref{fig:KekCirc}(a), therein, we have considered $E_0=9$ V/m and $\Omega=0.5$ THz, with the Kekul\'e parameter $\Delta_0=0.1$. It can be seen that two different quasienergy band gaps appear at the Dirac point, one for the external cones (shown schematically in green in Fig. \ref{fig:KekCirc}(b))

\begin{equation}
    \varepsilon_{\mathrm{Y},-}^{g}=\alpha-\beta=  \hbar\Omega \left(\sqrt{1+(2\tilde{E})^2}-\sqrt{1+(2\Delta_0\tilde{E})^2}
    \right),
    \label{eq:epsilonplus}
\end{equation}
and another one for the internal cones (shown schematically in blue in Fig. \ref{fig:KekCirc}(b))
\begin{equation}
   \varepsilon_{\mathrm{Y},+}^{g}=\varepsilon_{\mathrm{Y},-}^{g}+2 \delta\varepsilon^{g}_{\mathrm{Y}}.
    \label{eq:cuasiecpkekY}
\end{equation}
The gap between the concentric cones is obtained from 
\begin{equation}
    \delta\varepsilon^{g}_{\mathrm{Y}}=\beta-\hbar\Omega=\\
    \hbar\Omega\left(\sqrt{1+(2\Delta_0\tilde{E})^2}-1\right).
\end{equation}

The previous results have a simple interpretation. Let us expand up to first order in  $\tilde{E}$ to find,
\begin{equation}
    \varepsilon^{g}_{\mathrm{Y},\pm} \approx
    2\hbar\Omega (1\pm\Delta_0^2)\tilde{E}^2.
    \label{eq:epsilonYpm}
\end{equation}

Now consider the pristine graphene case $\Delta_0=0$ which results in $\delta\varepsilon^{g}_{\mathrm{Y}}=0$ and a gap $\varepsilon_{\mathrm{Y},\pm}^{g}=2\hbar\Omega \tilde{E}^2$. This represents a transition from a valence state at quasienergy $-\varepsilon^{g}_{\mathrm{Y}}/2$ to a final state in the conduction band with quasienergy $\varepsilon^{g}_{\mathrm{Y}}/2$. The transition is induced by resonant photon absorption of energy $2\hbar\Omega$ in the very weak field case, obtainable also with usual perturbation techniques\cite{KuboKek}. Higher order terms $\tilde{E}$  given by the Floquet theory are a dressing of the transition. As expected, such dressing is small yet is vital 
in Floquet  theory as otherwise the solution lies in a gap and thus is unstable. By turning on the $\Delta_0$ Kekul\'e
ordering parameter, we have a small detuning due to the  spatial modulation. This produces satellite peaks around each resonant frequency of the non-modulated system, a phenomena akin to beating in classical physics\cite{phasons,satija}.

\subsubsection{Kek-O texture under circularly polarized radiation}

To study the dynamical spectrum of charge carriers in Kek-O ($\nu=0$) textured graphene under circularly polarized light, we proceed as before (see Appendix \ref{Append-A}), we consider a weak electromagnetic field and obtain the following gapped quasienergy spectrum, 
\begin{equation}\label{QuasiOCirc}
    \varepsilon_{\mathrm{O}}^{\xi,\eta}(k)=\eta\frac{1}{2}\sqrt{(6t_0\Delta_0-\xi\alpha+\xi\hbar\Omega)^2+(2v\hbar k)^2}.
\end{equation}
We can see that the there is no degeneration when compared with the spectrum in the absence of external field. The quasienergy spectrum now consists of two concentric cones with different gaps at the Dirac point. For the external cones we find a gap given by
\begin{equation}
    \varepsilon_{\mathrm{O},-}^g=6t_0\Delta_0-\alpha+\hbar\Omega=6t_0\Delta_0+\hbar\Omega\left(1-\sqrt{1+(2\tilde{E})^2}\right),
    \label{eq:kekoplus}
\end{equation}
and for the internal cones
\begin{equation}
     \varepsilon_{\mathrm{O},+}^g=6t_0\Delta_0+\alpha-\hbar\Omega=6t_0\Delta_0-\hbar\Omega\left(1-\sqrt{1+(2\tilde{E})^2}\right),
      \label{eq:kekominus}
\end{equation}
with the gap between the concentric cones given by
\begin{equation}
    \delta\varepsilon_{\mathrm{O}}^g=\alpha-\hbar\Omega.
\end{equation}
as shown schematically in Fig. \ref{fig:KekCirc}(d).

The quasienergy spectrum for the $\nu=0$ Kek-O texture under circularly polarized radiation in the weak electromagnetic field regime \eqref{aproxx} is shown in Fig. \ref{fig:KekCirc}(c), where we have considered $E_0=9$ V/m and $\Omega=0.5$ THz, with the Kekul\'e parameter $\Delta_0=0.1$.

From Eqs. (\ref{eq:kekoplus}) and 
(\ref{eq:kekominus}), we obtain that,

\begin{equation}
    \varepsilon_{\mathrm{O},\pm}^g \approx 6t_0\Delta_0 \pm (2\hbar\Omega) \tilde{E}^2.
\end{equation}

Up to order zero in $\tilde{E}$ the result is just the same static gap $6t_0\Delta_0$  already present in the system without radiation. Then we have the transition from the valence to conduction band induced by the photon with energy $2\hbar\Omega$ as in Eq. (\ref{eq:epsilonYpm}).

\subsection{Linearly polarized light}

Considering normal incidence of a linearly polarized electromagnetic wave defined by the vector potential
\begin{equation}\label{vectPotLin}
    \boldsymbol{A}=\frac{E_0}{\Omega}\big(\cos{(\Omega t)},\:0\big),
\end{equation}
where for simplicity we consider the polarization along the $\hat{\boldsymbol{x}}$ direction, and again we have neglected the third dimension. The corresponding electric field is given by $\boldsymbol{E}=-\partial\boldsymbol{A}/\partial t=-E_0\big(\sin(\Omega t),\:0\big)$. In the next subsections we show the resulting spectrum for the two kinds of Kekul\'e bond texture.

\subsubsection{Kek-Y texture under linearly polarized light}
Consider a Kek-Y textured graphene irradiated with linearly polarized light, again we take $\nu=1$ and a real $\tilde{\Delta}=\Delta_0$. For this case, we are interested in the high frequency regime, such that the energy of a photon is larger than the energy of charge carriers in pristine graphene,
\begin{equation}\label{highFreq}
    v\hbar k\ll\hbar\Omega.
\end{equation} 
We can analytically solve the Dirac Eq. \eqref{DiracEquation} under the last considerations (see Appendix \ref{Append-B}) and obtain the  following gapless quasienergy spectrum 
\begin{eqnarray}\label{QuasiYLin}
    \nonumber\varepsilon_{\mathrm{Y}}^{\xi,\eta}(\boldsymbol{k})=\eta v\hbar k\Big(&&\sqrt{\cos^2{(\theta)}+J_0^2(2\tilde{E})\sin^2{(\theta)}}\\
   \nonumber &&+\:\xi \Delta_0\sqrt{\cos^2{(\theta)}+J_0^2(2\Delta_0\tilde{E})\sin^2{(\theta)}}\Big),\\
\end{eqnarray}
where $J_0(z)$ is the Bessel function of the first kind. \\

\begin{figure}[t]
\centering
\large(a)\qquad\qquad\qquad\qquad\qquad\large(b)\\
\includegraphics[scale=0.32]{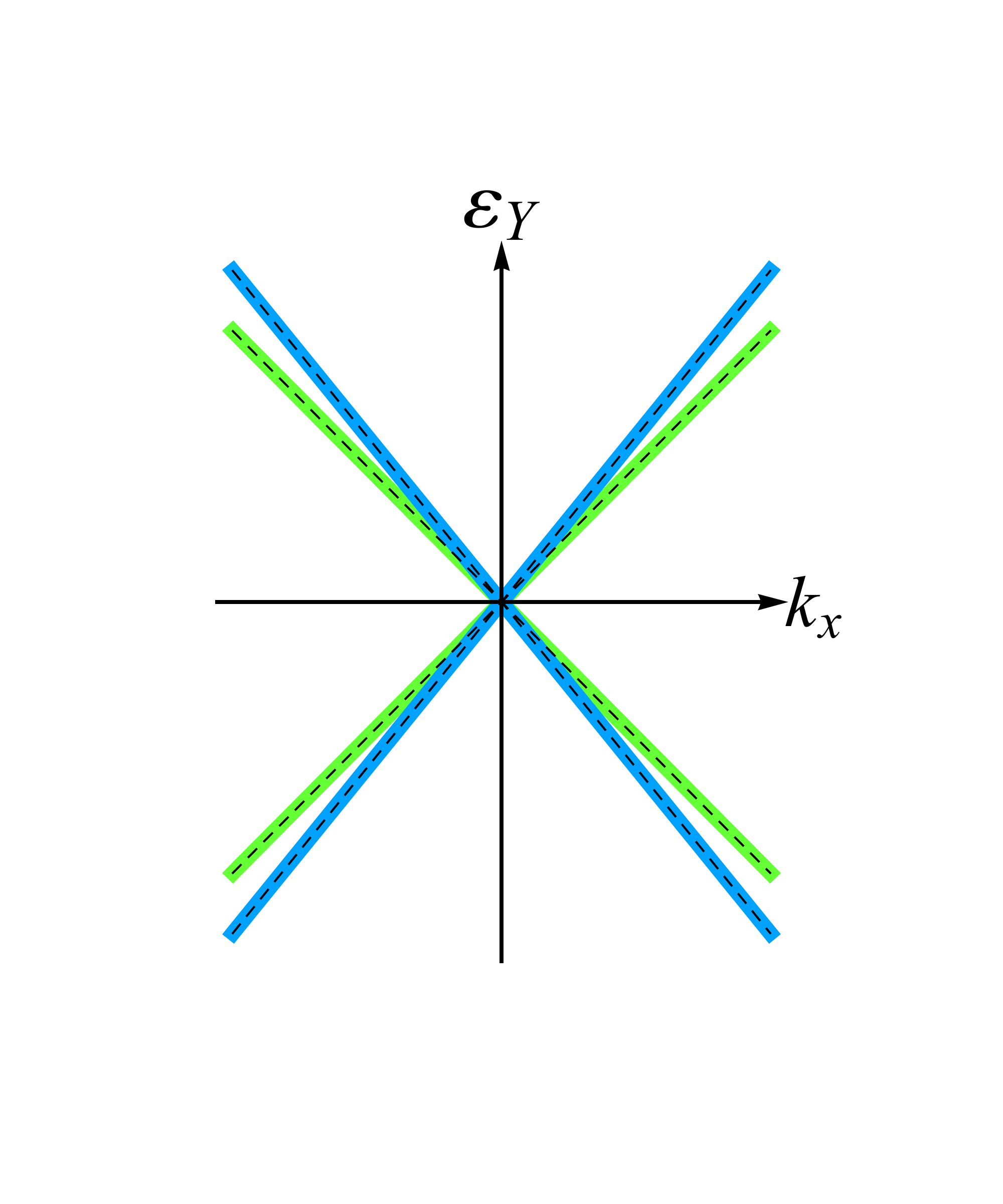}
\includegraphics[scale=0.32]{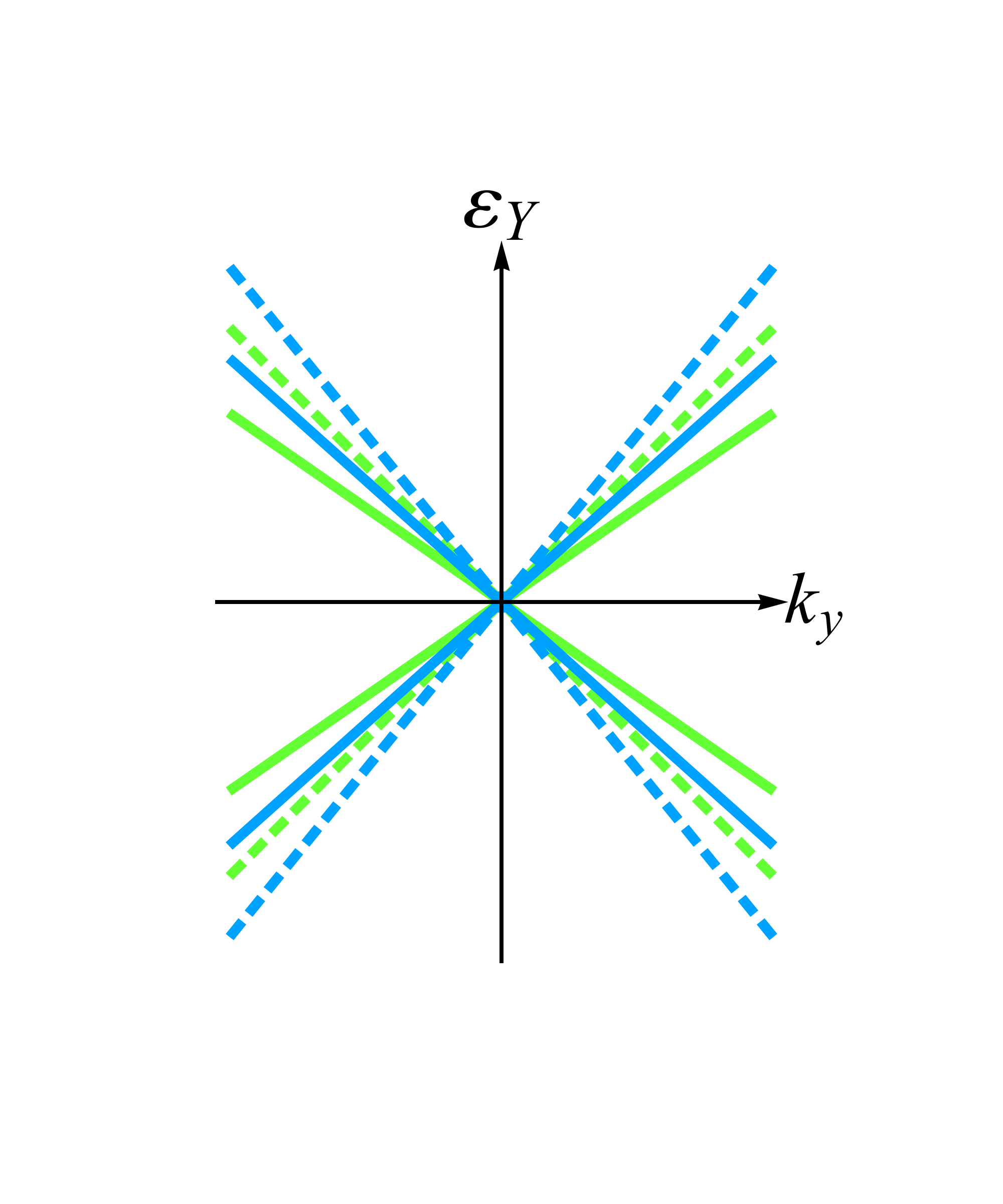}

\large(c)\qquad\qquad\qquad\qquad\qquad\large(d)\\
\includegraphics[scale=0.31]{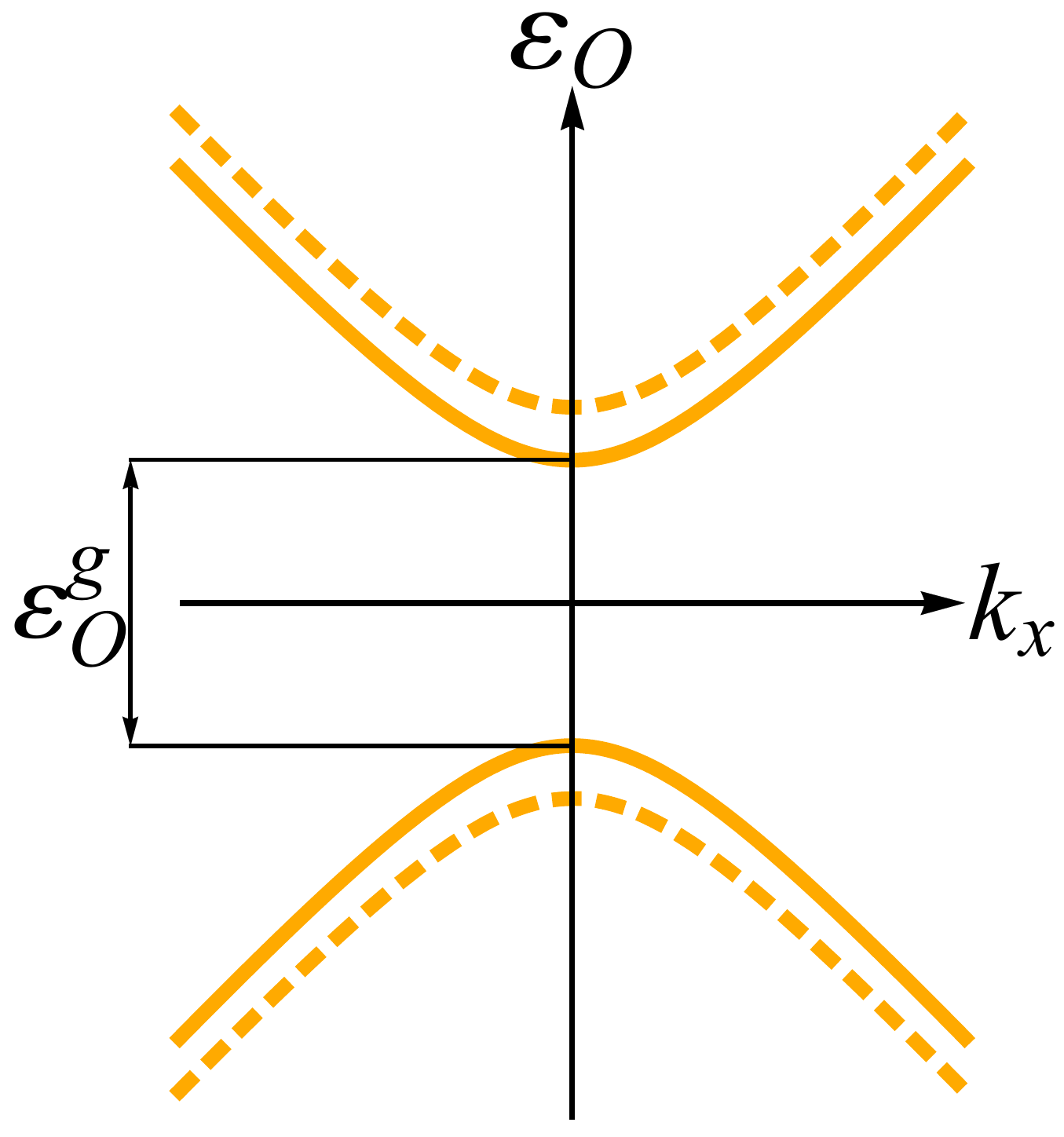}
\includegraphics[scale=0.31]{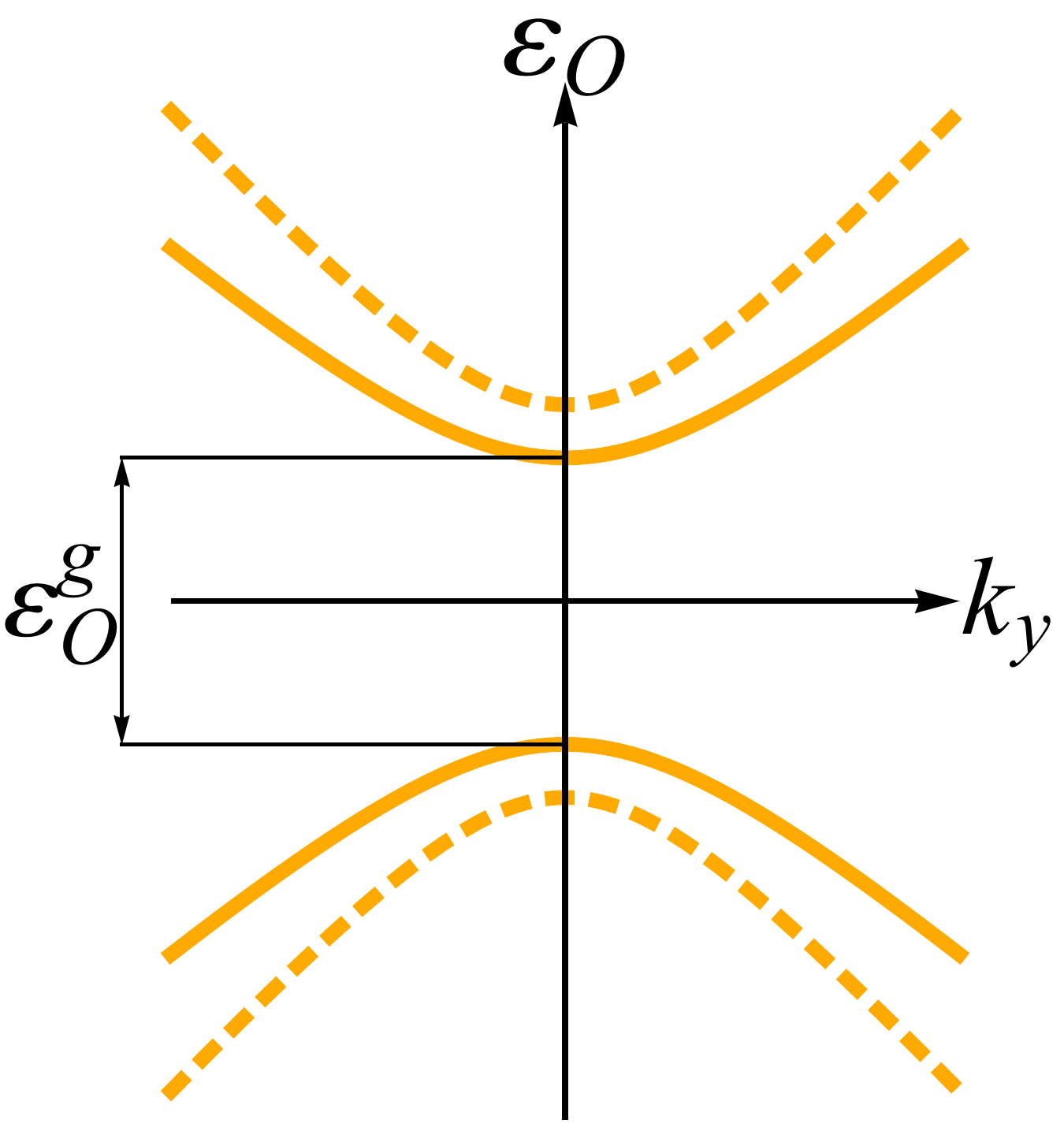}
\caption{\label{fig:KekLin}
Schemes of the quasienergy spectrum of Kek-Y distorted graphene with a cut along (a) the $k_x$ direction, and (b) the $k_y$ direction, irradiated with a linearly polarized electromagnetic wave in the high frequency regime \eqref{highFreq}. Similarly, it is shown a cut of the quasispectrum of Kek-O distorted graphene
along (c) the $k_x$ direction, and (d) the $k_y$ direction. Here, the dotted lines represents the energy bands in absence of the external wave. It is shown how the spectrum is modified perpendicular to the direction of polarization, and how the natural energy gap in the Kek-O distorted graphene is reduced.}
\end{figure}

Fig.~\ref{fig:KekLin}(a) shows a cut of the spectrum along the $k_x$ direction. As can be seen from Eq. \eqref{QuasiYLin}, for this parallel direction there is no change in the spectrum; this holds even without taking the high frequency approximation as for $\theta=0$ we obtain $\varepsilon_{\mathrm{Y}}^{\xi,\eta}(\boldsymbol{k})=E_{\mathrm{Y}}^{\xi,\eta}(k)$. This means that transitions are not induced by the external field as for this kind of light the symmetry is not broken. 
For any other direction of momentum, the application of the linearly polarized light results in a direction-dependent Fermi velocity, as shown schematically in Fig.~\ref{fig:KekLin}(b). 

The change in the direction-dependent Fermi velocity can be obtained by developing the square root in Eq. (\ref{QuasiYLin}),
\begin{eqnarray}
    \varepsilon_{\mathrm{Y}}^{\xi,\eta}(\boldsymbol{k})\approx E_{\mathrm{Y}}^{\xi,\eta}(k)+ \eta \Delta v(\theta) \hbar k,
\end{eqnarray}
where,
\begin{equation}
    \Delta v(\theta)=\left[\frac{J_0^2(2\tilde{E})+\xi \Delta_0 J_0^2(2\Delta_0\tilde{E})}{2}-1\right]\sin ^2(\theta).
\end{equation}
Notice that for linearly polarized light, we do not need to assume  $\tilde{E}\ll 1$, thus in Eq. (\ref{QuasiYLin}) the system can reach the condition $J_0(2\tilde{E})=0$ or $J_0(2\Delta_0 \tilde{E})=0$. A zero of the Bessel function will imply a band nearly flat in the direction $\theta=\pm\pi/2$.

\subsubsection{Kek-O texture under linearly polarized light}

Finally, for Kek-O ($\nu=0$) textured graphene under linearly polarized light under the high frequency regime, we found the following degenerate gapped quasienergy spectrum 
\begin{eqnarray}\label{QuasiOLin}
    \nonumber\varepsilon_{\mathrm{O}}^{\eta}(\boldsymbol{k})=\eta&&\Big\{(v\hbar k)^2\big[\cos^2{(\theta)}+J_0^2(2\tilde{E})\sin^2{(\theta)}\big]\\
    &&+\:(3t_0\Delta_0)^2J_0^2(2\tilde{E})\Big\}^{1/2}.
\end{eqnarray}
We note that for this case, the incidence of radiation is equivalent to perform $k_y\rightarrow |J_0(2\tilde{E})|k_y$, and to modify the band gap by a factor, $\Delta_0\rightarrow |J_0(2\tilde{E})|\Delta_0$, such that the quasienergy band gap is $\varepsilon_{\mathrm{O}}^g=6t_0\Delta_0|J_0(2\tilde{E})|$. The quasienergy spectrum for Kek-O textured graphene under linearly polarized radiation is shown schematically in Figs.~\ref{fig:KekLin}(c)-(d).
Whenever $J_0(2\tilde{E})=0$, the quasispectrum becomes gapless, then we have that, 
\begin{equation}\label{QuasiOLin2}
    \varepsilon_{\mathrm{O}}^{\eta}(\boldsymbol{k})=\eta v \hbar k\cos\theta =\eta v  \hbar k_x.
\end{equation}
This also shows that for $\theta=\pm\pi/2$ a non-dispersive band is observed. Which means that electrons are localized in the $y$ direction. Hence, we can find the value of the electric field to obtain this non-dispersive band. If we take into account the first root of the Bessel function $J_{0}(2\tilde{E})$, this condition implies that $2\tilde{E}\approx 2.405$, and therefore using Eq. \eqref{aproxx} with a high-frequency $\Omega=1713$ THz, we find $E_0\approx 2.32$ V/nm. 
These values of intensity and frequency for the electromagnetic field can be achieved, for example, using a Ti:sapphire laser (650 - 1100 nm) with a power per unit area of $7.16\times 10^{-3}\,$W/nm$^2$, as in recent graphene photocurrent experiments\cite{higuchi2017light,heide2018coherent}.

\section{dc conductivity}\label{sec:Sec4}

As is now well established, light changes a 2D material conductivity\cite{AnnualReview-Oka}. It is interesting to explore such photoconductivity effect for the studied system. 
As an example we will calculate the dc conductivity of the  Kek-Y ($\nu=1$) distorted graphene under circularly polarized radiation. This calculation is made using the Boltzmann transport theory \cite{Mahan1990,Rossiter1991}
which requires the introduction of a relaxation mechanism. Here is taken as random-distributed delta function scatters such that the scattering potential is written as follows\cite{Yudin2016}
\begin{equation}
   U(\boldsymbol{r})=\sum_{j=1}^{N}U_0\delta(\boldsymbol{r}-\boldsymbol{r}_j),
\end{equation}
with $\boldsymbol{r}$ the position vector. It is important to remark that in Kekul\'e-distorted graphene the Brillouin zone is folded and the two Dirac cones are brought into the center, as mentioned before. 
Consequently, electronic transitions between states in the two valleys are now possible (inter-valley transport), in addition to those between states in a single valley (intra-valley transport).
The probability of horizontal transitions of conduction electrons ($\eta=+$) between the cone $\xi$ with wave vector $\boldsymbol{k}$, and the cone $\xi'$ with wave vector $\boldsymbol{k}'$, per unit time, is calculated using the Born scattering expression\cite{LanduBookQuantum,kibis}, and has the following form
\begin{equation}
    w_{\boldsymbol{k}'\boldsymbol{k}}^{\xi',\xi}=\frac{2\pi}{\hbar}|\chi_{\boldsymbol{k}'\boldsymbol{k}}^{\xi',\xi}|^2|U_{\boldsymbol{k}'\boldsymbol{k}}|^2\delta\big(\varepsilon_{\mathrm{Y}}^{\xi',+}(k')-\varepsilon_{\mathrm{Y}}^{\xi,+}(k)\big),
\end{equation}
where 
\begin{equation}
    \chi_{\boldsymbol{k}'\boldsymbol{k}}^{\xi',\xi}=\sum_{n=1}^{4}b_{n}^{\xi,+}(\boldsymbol{k})\big(b_{n}^{\xi',+}(\boldsymbol{k}')\big)^{*},
\end{equation}
with each $b_{n}^{\xi,+}(\boldsymbol{k})$ given in Appendix \ref{Append-A}, and the square modulus of the matrix elements of the scattering potential\cite{Yudin2016} is given by
\begin{equation}\label{ScatPot}
    |U_{\boldsymbol{k}'\boldsymbol{k}}|^2=\Bigg|\frac{1}{A}\int_Ad^{2}\boldsymbol{r} \:e^{i(\boldsymbol{k}-\boldsymbol{k}')\cdot\boldsymbol{r}}U(\boldsymbol{r})\Bigg|^2=\frac{N_{A}}{A}U_0^2,
\end{equation}
where $A$ is the sample area and $N_A=N/A$ is the density of impurities.

\begin{figure}[t]
    \centering{\large(a)}\\
    \includegraphics[scale=0.455]{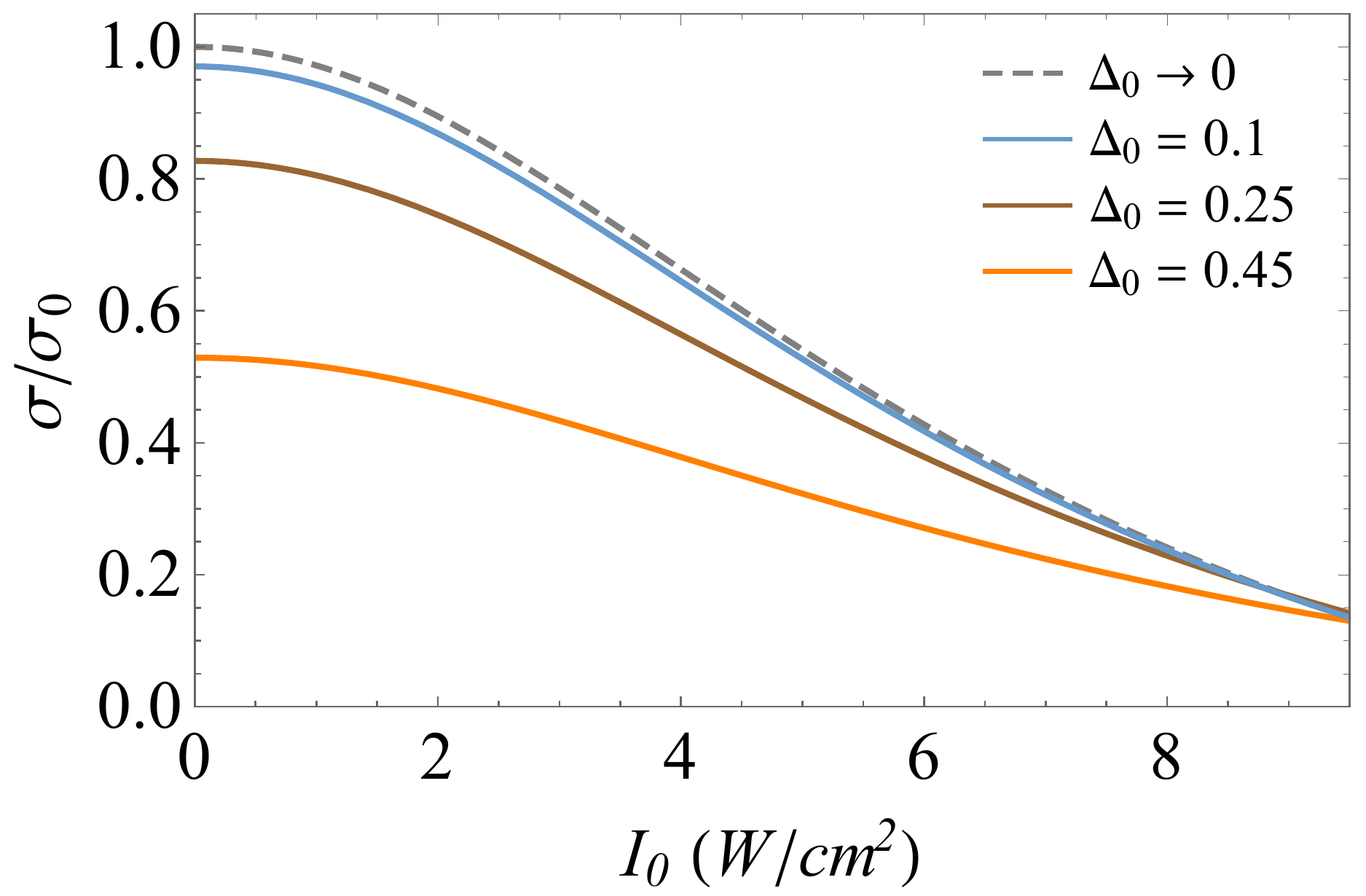}
     \centering{\large(b)}\\
    \includegraphics[scale=0.455]{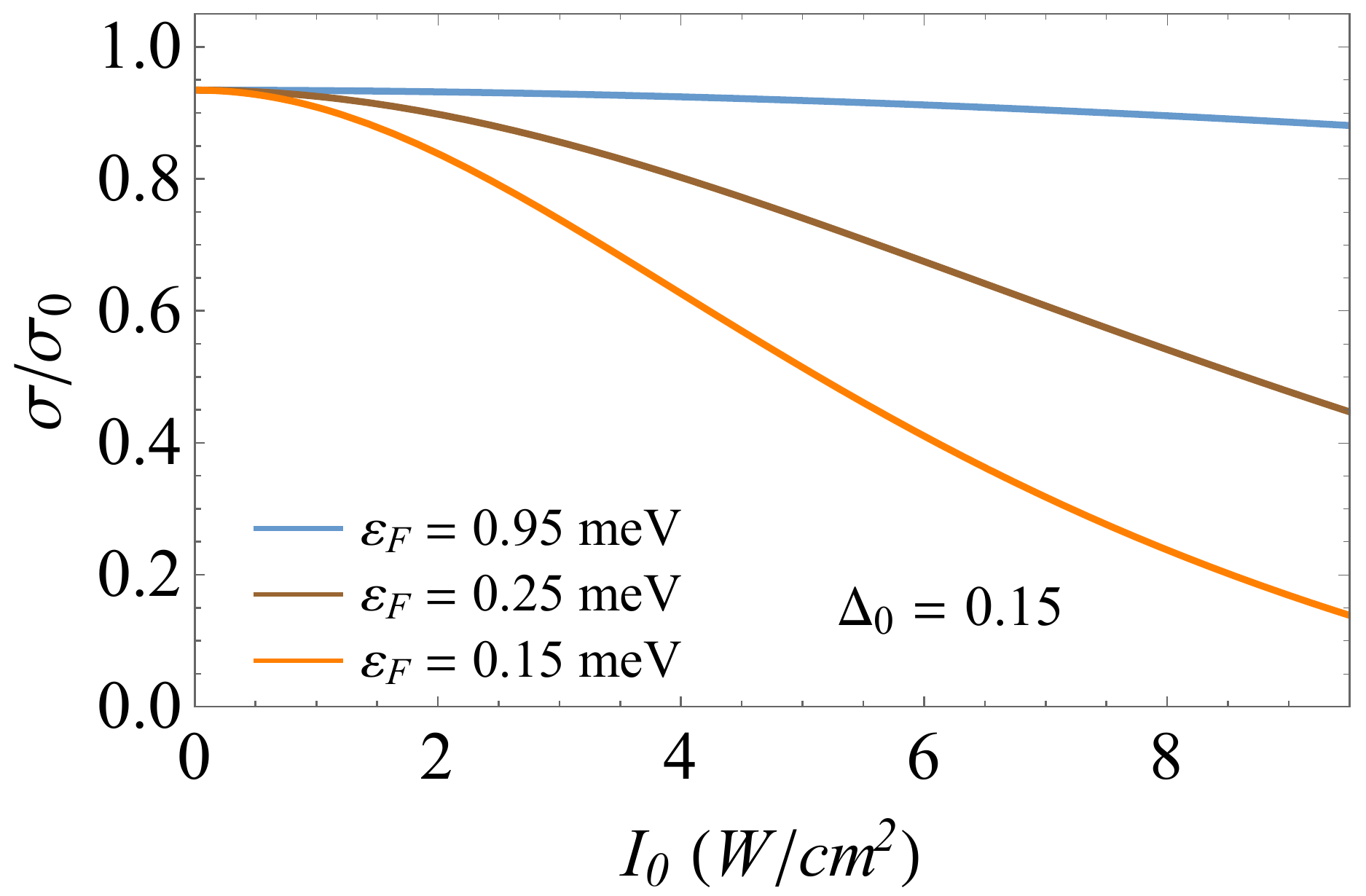}
    \caption{\label{fig:cond}dc conductivity of the Kek-Y distorted graphene under circularly polarized radiation as a function of the irradiance $I_0$ (where $I_0=\epsilon_0E_0^2c/2$).  (a) shows the behavior of the dc conductivity for different values of the Kekul\'e parameter $\Delta_0$ with $\varepsilon_F=0.15\,$ meV. Notice how in the limit of $\Delta_0\rightarrow0$  we recover the dc conductivity of graphene \cite{kibis}. In (b), we show the behavior of the conductivity for different values of the Fermi energy and $\Delta_0=0.15$. In both panels, $\sigma_0$ ($4\hbar e^2v^2/\pi N_AU_0^2$) is the dc conductivity in graphene when $I_0=0$. The photon energy of the polarized wave has a value  $\hbar\Omega=4\,$meV.}
\end{figure}
\begin{figure}[t]
    \centering{\large(a)}\\
    \includegraphics[scale=0.645]{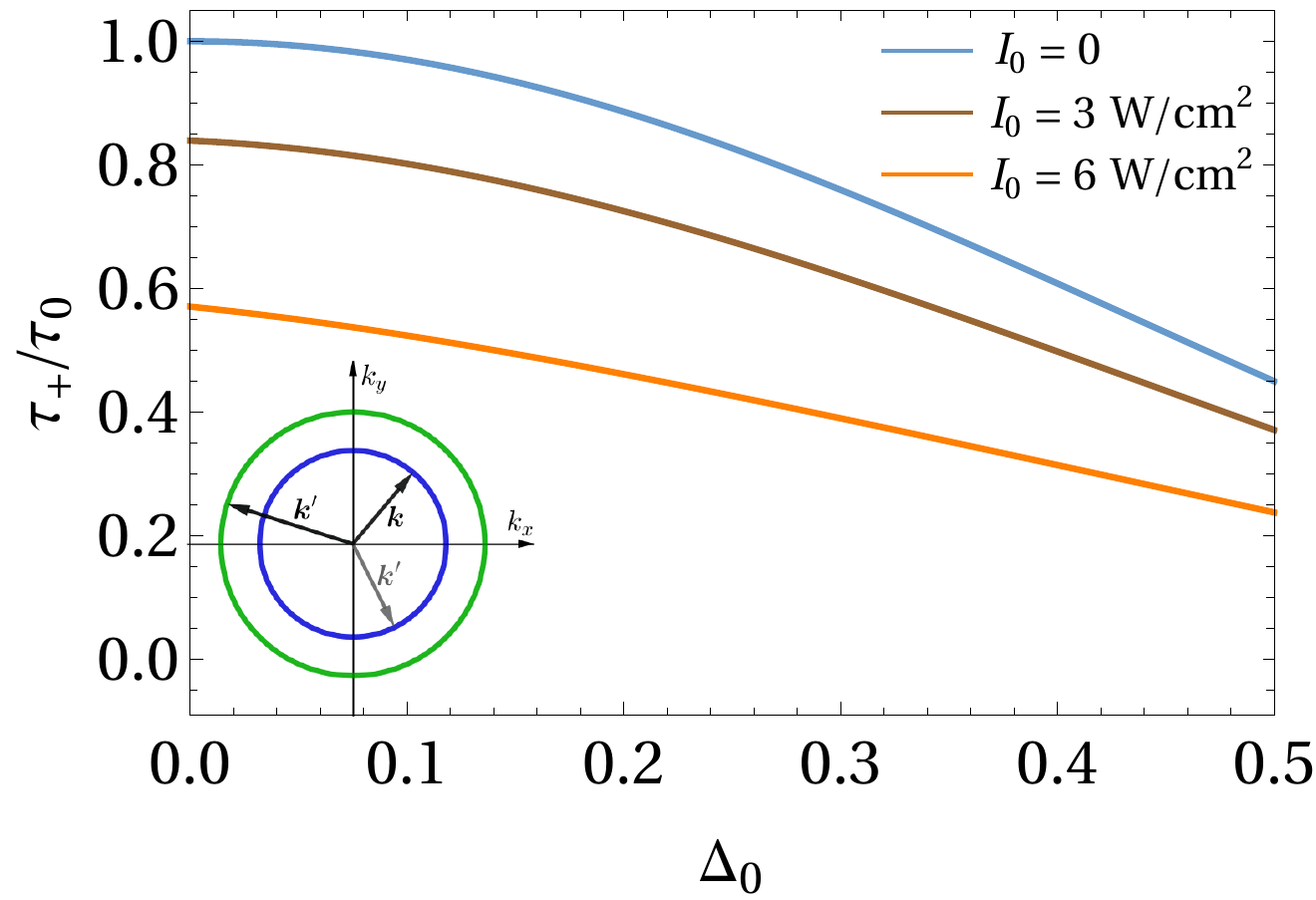}
     \centering{\large(b)}\\
    \includegraphics[scale=0.645]{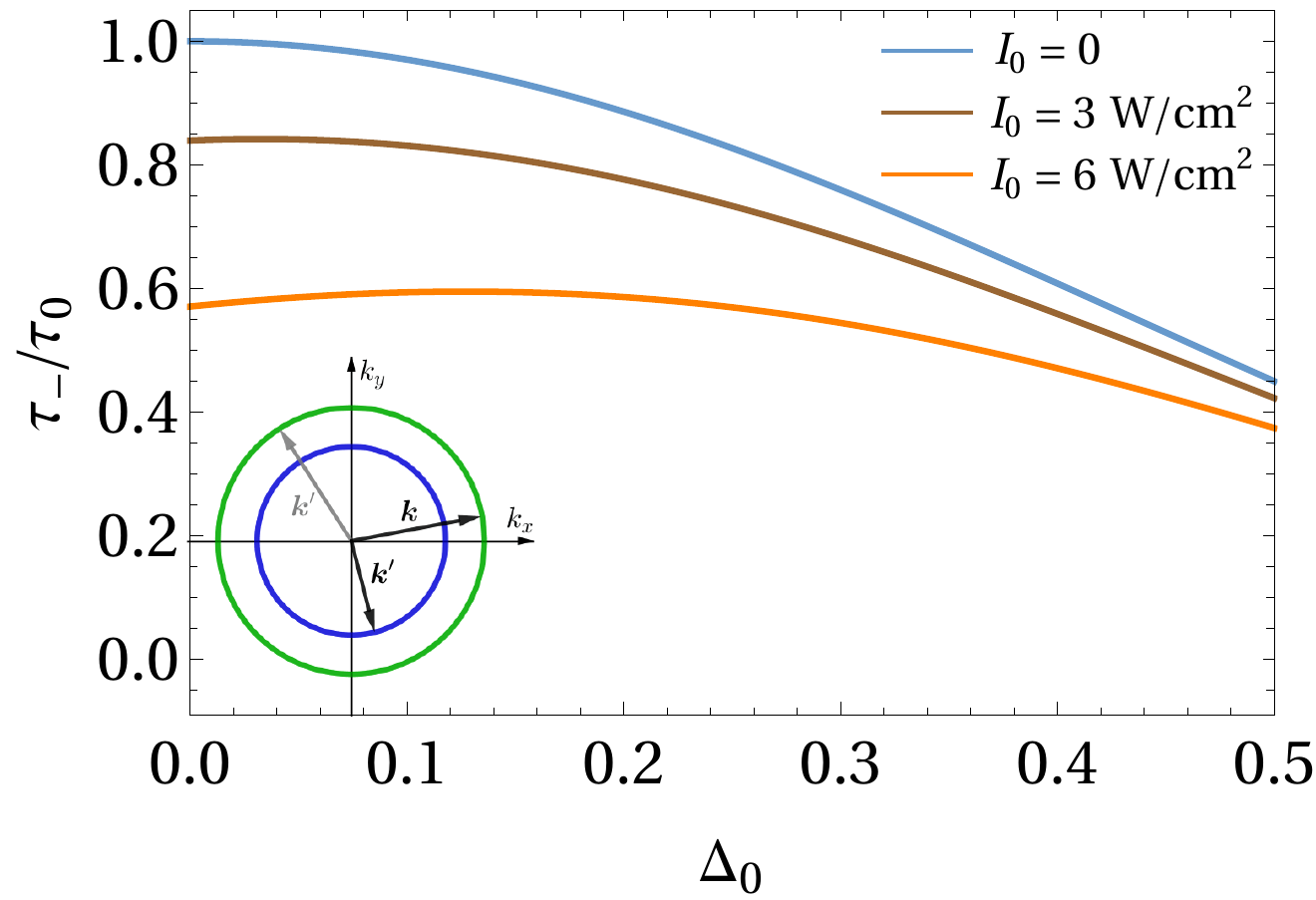}
    \caption{\label{fig:times}Relaxation time as a function of the Kekul\'e parameter $\Delta_0$ for different values of the irradiance $I_0$ ($\epsilon_0E_0^2c/2$) and  $\varepsilon_F=0.15$. (a) shows the relaxation time for transitions with the internal cone as the initial state ($\xi=+$). The inset shows a horizontal cut in the quasispectrum to illustrate two possible horizontal transitions. In (b), we show the relaxation time with the external cone as the initial state ($\xi=-$). In the same way,  the inset shows the two possibles horizontal transitions. In both panels, we use $\tau_0=4\hbar^3v^2/\varepsilon_FN_AU_0^2$ which corresponds to the pristine graphene relaxation time when $I_0=0$.}
\end{figure}

If we apply a stationary electric field $\boldsymbol{E}$ to the sample, the Boltzmann equation for the current density at zero temperature and a Fermi energy $\varepsilon_F$ in the conduction band is given by
\begin{equation}\label{boltzmann}
    \boldsymbol{J}=\sum_{\xi}\frac{e^2}{(2\pi)^2}\int d^2\boldsymbol{k}\big(\boldsymbol{E}\cdot\boldsymbol{v}_{\xi}(k)\big)\tau_{\xi}(\boldsymbol{k})\boldsymbol{v}_{\xi}(k)\delta\big(\varepsilon_{\mathrm{Y}}^{\xi,+}(k)-\varepsilon_F\big),
\end{equation}
where $\boldsymbol{v}_{\xi}(k)=(1/\hbar)\nabla_{\boldsymbol{k}}\varepsilon_{\mathrm{Y}}^{\xi,+}(k)$ is the group velocity and $\tau_{\xi}(\boldsymbol{k})$ is the relaxation time, which in the more general case is written as the following integral equation \cite{relaxTime}
\begin{equation}\label{time}
    \frac{1}{\tau_{\xi}(\boldsymbol{k})}=\sum_{\xi'}A\int d^2\boldsymbol{k}\left[1-\frac{\tau_{\xi'}({\boldsymbol{k'})}\big(\boldsymbol{E}\cdot\boldsymbol{v}_{\xi'}(k')\big)}{\tau_{\xi}(\boldsymbol{k})\big(\boldsymbol{E}\cdot\boldsymbol{v}_{\xi}(k)\big)}\right]w_{\boldsymbol{k}'\boldsymbol{k}}^{\xi',\xi}.
\end{equation}
The sum in Eq. \eqref{boltzmann} runs over the initial states, and the sum in Eq. \eqref{time} runs over the final states. Given the azimuth symmetry, we consider an electric field in the $\hat{\boldsymbol{x}}$ direction $\boldsymbol{E}=(E_x,\:0)$, such that the conductivity will be given by $\sigma=J_x/E_x$. The subsequent steps to calculate the current density as well as the relaxation time are shown in Appendix \ref{Append-C}. It is found that the dc conductivity has the following form
\begin{equation}\label{finalsigma}
\sigma=\sum_{\xi}\frac{e^2}{2\pi\hbar} k_F^{\xi}\tau_{\xi}(k_F^{\xi})v_{\xi}(k_F^{\xi}),
\end{equation}
where $k_F^{\xi}$ is the Fermi wave vector such that $\varepsilon_{\mathrm{Y}}^{\xi,+}(k_F^{\xi})=\varepsilon_F$, and $v_{\xi}(k)=|\boldsymbol{v}_{\xi}(k)|$. Also, we have considered the spin degeneracy (factor 2).  In Fig. \ref{fig:cond}(a) we show the conductivity at different values of $\Delta_0$ and the graphene limit $\Delta_0\rightarrow0$, recovering previous results\cite{kibis}. Also, in Fig. \ref{fig:cond}(b) we show the conductivity with a fixed Kekul\'e parameter $\Delta_0$ for different values of the Fermi energy. As expected, Fig. \ref{fig:cond}(a) shows that the Kekul\'e bond pattern decreases $\sigma$. Moreover, in the limit $\Delta_0 \rightarrow 1$, the conductivity goes to zero. In this limit, the hopping amplitude of the longer bonds in Fig. \ref{fig:KekO_KekY} (a) is zero resulting in a disconnected lattice of red Y bond patterns. Therefore the conductivity is zero.

By looking at Eq. \eqref{finalsigma}, is clear that $\sigma$ is modified mainly by the new relaxation scattering channels, a typique  result for modulated systems \cite{Oliva_Leyva_2014,Taboada_2014}. Notice that in Eq. \eqref{finalsigma} two kinds of relaxation appear. In one kind, the initial state is the internal cone. Then the electron can be scattered into a state in the same cone or into a state in a different cone. These transitions are sketched out by the cuts of the energy dispersion seen in Fig. \ref{fig:times} (a). For pristine graphene, the relaxation between different cones is forbidden while here is allowed. In the other kind, the initial state is in the external cone. The corresponding relaxation times are given in Eqs. \eqref{tauPlus} and \eqref{tauMinus} for a fixed Fermi energy. In Fig. \ref{fig:times} we show the relaxation time as a function of the Kekul\'e parameter $\Delta_0$ for different values of the irradiance of the external electromagnetic wave. Observe that for $I_0=0$, the relaxation time decreases as a function of $\Delta_0$ explaining the results of Fig. \ref{fig:KekO_KekY}.

\section{Conclusions}\label{sec:Sec5}
We studied the effects of normal electromagnetic radiation on Kekul\'e distorted graphene. We presented analytical expressions for the quasienergy spectrum for both textures, Kek-Y and Kek-O, and considering two types of polarization, circular and linear. Circularly polarized radiation opens a gap in the otherwise gapless spectrum of Kek-Y distorted graphene, while it breaks the valley degeneracy in the gapped spectrum of Kek-O textured graphene. To further characterize these gaped systems we calculated, by using the Boltzmann approach, the dc current as a function of the field intensity, considering inter- and intra-valley contributions.
We found that the total conductivity is decreased by the Kekul\'e distortion as relaxation times are decreased due to the new open scattering channels, as for example transitions between the internal and external cones. On the other hand, linearly polarized radiation does not open a gap in the Kek-Y distorted graphene, but modify the gap in Kek-O graphene. Moreover, it breaks the angular symmetry in the quasienergy spectrum for both textures. An interesting result is that for linear polarization, non-dispersive bands can appear by a precise tuning of the light field parameters, thus inducing dynamical localization.
In all cases we successfully  recover the expressions for pristine irradiated graphene \cite{kibis} in the limit $\Delta_0\rightarrow0$.

\section{Acknowledgements}\label{sec:Acknowledgements}

We thank UNAM DGAPA-PROJECT IN102620. V.G.I.S and J.C.S.S. acknowledge the total support from DGAPA-UNAM fellowship. M.A.M. and R.C.-B. thank Jesús Maytorena for discussions on this paper.

\onecolumngrid

\appendix{}
\section{Circularly polarized light}\label{Append-A}

We detail below the procedure to obtain the quasienergy spectrum and the corresponding spinors of Kek-Y ($\nu=1$) distorted graphene under circularly polarized light. We start from the Eq. \eqref{NonStatDiracE} expressed as
\begin{eqnarray}\label{TimeDepenKekY}
    i\hbar\frac{d}{dt}\boldsymbol{\psi}(t)=-ev\Big[&&\big(\boldsymbol{A}\cdot\boldsymbol{\sigma}\big)\otimes\mu_0
   +\Delta_0\sigma_0\otimes\big(\boldsymbol{A}\cdot\boldsymbol{\mu}\big)\Big]\boldsymbol{\psi}(t),
\end{eqnarray}
where $\boldsymbol{A}$ is given by Eq. \eqref{potential} and with an extra set of Pauli matrices $\mu_{i}$ acting on the valley degree of freedom. To solve the previous equation, we make the following ansatz
\begin{equation}\label{ansatzKekY}
\boldsymbol{\psi}(t)=e^{-i\epsilon t/\hbar}U(t)\boldsymbol{\varphi},
\end{equation}
where $\epsilon$ is a characteristic exponent,  $\boldsymbol{\varphi}$ is a four-component spinor and $U(t)$ is a unitary transformation given by
\begin{equation}
U(t)=\begin{pmatrix}
e^{-i\Omega t}&0&0&0\\
0&1&0&0\\
0&0&1&0\\
0&0&0&e^{i\Omega t}
\end{pmatrix}.
\end{equation}
Substituting Eq. \eqref{ansatzKekY} into Eq. \eqref{TimeDepenKekY} (with a real  $\tilde{\Delta}=\Delta_0$) we find 

\begin{equation}\label{EigenKekY}
    -\hbar\Omega\begin{pmatrix}
    1&\tilde{E}&\Delta_0\tilde{E} &0\\
    \tilde{E}&0&0&\Delta_0\tilde{E} \\
    \Delta_0\tilde{E} &0&0&\tilde{E}\\
    0& \Delta_0\tilde{E} &\tilde{E}&-1\\
    \end{pmatrix}
    \boldsymbol{\varphi}= \epsilon\:\boldsymbol{\varphi},
\end{equation}
where $\tilde{E}=eE_0v/\hbar\Omega^2$.

Analyzing the eigenvalue problem of Eq. \eqref{EigenKekY} we find $\epsilon$ and $\boldsymbol{\varphi}$. Hence, Eq. \eqref{ansatzKekY} is expressed as

\begin{equation}\label{ketsKekY}
\boldsymbol{\psi}_{n}(t)=\boldsymbol{\psi}_{\xi,\eta}(t)=
\frac{e^{i\eta(\alpha+\xi\beta)t/2\hbar}}{2\sqrt{\alpha\beta}}\begin{pmatrix}
\xi e^{-i\Omega t}\sqrt{(\alpha+\eta\hbar\Omega)(\beta+\xi\eta\hbar\Omega)}\\ \xi\eta\sqrt{(\alpha-\eta\hbar\Omega)(\beta+\xi\eta\hbar\Omega)}\\
\eta\sqrt{(\alpha+\eta\hbar\Omega)(\beta-\xi\eta\hbar\Omega)}\\
e^{i\Omega t}\sqrt{(\alpha-\eta\hbar\Omega)(\beta-\xi\eta\hbar\Omega)}
\end{pmatrix},
\end{equation}
with $\xi=\pm$, $\eta=\pm$, $n\in\{1,2,3,4\}$, according to the prescription $\boldsymbol{\psi}_1(t)=\boldsymbol{\psi}_{+,+}(t)$, $\boldsymbol{\psi}_2(t)=\boldsymbol{\psi}_{-,-}(t)$, $\boldsymbol{\psi}_3(t)=\boldsymbol{\psi}_{-,+}(t)$ and $\boldsymbol{\psi}_4(t)=\boldsymbol{\psi}_{+,-}(t)$; and we have defined the following constants
\begin{equation}
\alpha=\hbar\Omega\sqrt{(2\tilde{E})^2+1}, \,\,\,\,\,\,\,\,\,\,  \beta=\hbar\Omega\sqrt{(2\Delta_0\tilde{E})^2+1}.  
\end{equation}



Now, keeping in mind Eq. \eqref{ketsKekY} in Eq. \eqref{ansatz}, the Dirac equation   \eqref{DiracEquation} for each coefficient $a_{m}(\boldsymbol{k},t)$ is
\begin{equation}\label{coef}
    i\hbar\frac{d}{dt}a_{m}(\boldsymbol{k},t)=\sum_{n=1}^{4}\Big[\boldsymbol{\psi}_m^{\dagger}(t)H_0\boldsymbol{\psi}_n(t)\Big] a_{n}(\boldsymbol{k},t)-\varepsilon a_{m}(\boldsymbol{k},t),\qquad m\in\{1,2,3,4\}.
\end{equation}

Here we focus on the weak electric field regime, therefore we thus assume that $\tilde{E}\ll1$, or alternatively $\alpha\approx\hbar\Omega$ and $\beta\approx\hbar\Omega$. Under this condition the Eq. \eqref{coef} is reduced as
\begin{equation}\label{eqCoeff}
i\hbar\frac{d}{dt}a_{m}(\boldsymbol{k},t)=\sum_{n=1}^{4}\mathbb{H}_{mn}(t)a_{n}(\boldsymbol{k},t)-\varepsilon a_{m}(\boldsymbol{k},t),
\end{equation}
where $\mathbb{H}_{mn}(t)\approx\boldsymbol{\psi}_m^{\dagger}(t)H_0\boldsymbol{\psi}_n(t)$ are the elements of the matrix
\begin{equation}
    \mathbb{H}(t)=v\hbar k\begin{pmatrix}
    0&e^{-i[(\alpha-\hbar\Omega)t/\hbar+\theta]}&\Delta_0 e^{-i[(\beta-\hbar\Omega)t/\hbar+\theta]}&0\\
    e^{i[(\alpha-\hbar\Omega)t/\hbar+\theta]}&0&0&\Delta_0 e^{-i[(\beta-\hbar\Omega)t/\hbar+\theta]}\\
    \Delta_0 e^{i[(\beta-\hbar\Omega)t/\hbar+\theta]}& 0&0&e^{-i[(\alpha-\hbar\Omega)t/\hbar+\theta]}\\
     0&\Delta_0 e^{i[(\beta-\hbar\Omega)t/\hbar+\theta]}&e^{i[(\alpha-\hbar\Omega)t/\hbar+\theta]}&0
    \end{pmatrix}.
\end{equation}

The next step is to find the quasienergy spectrum $\varepsilon$. Therefore, we assumed that each coefficient $a_{m}(\boldsymbol{k},t)$ in Eq. \eqref{eqCoeff} can be expressed as
\begin{eqnarray}\label{amExpansion}
a_{m}(\boldsymbol{k},t)=\sum_{n=1}^4U_{mn}(t)b_{n}(\boldsymbol{k}),    
\end{eqnarray}
where
\begin{equation}
U(t)=\begin{pmatrix}\label{UnitaryTransTwo}
e^{i(2\hbar\Omega-\alpha-\beta)t/2\hbar}&0&0&0\\
0&e^{i(\alpha-\beta)t/2\hbar}&0&0\\
0&0&e^{i(\beta-\alpha)t/2\hbar}&0\\
0&0&0&e^{i(\alpha+\beta-2\hbar\Omega)t/2\hbar}
\end{pmatrix},
\end{equation}
is a unitary transformation. Introducing Eq. \eqref{amExpansion} into \eqref{eqCoeff} we get the following eigenvalue problem
\begin{equation}
    \varepsilon\begin{pmatrix}
    b_{1}(\boldsymbol{k})\\
    b_{2}(\boldsymbol{k})\\
    b_{3}(\boldsymbol{k})\\
    b_{4}(\boldsymbol{k})
    \end{pmatrix}=\begin{pmatrix}
    \hbar\Omega-(\alpha+\beta)/2&v\hbar k e^{-i\theta}&v\hbar k\Delta_0  e^{-i\theta}&0\\
    v\hbar k e^{i\theta}&(\alpha-\beta)/2&0&v\hbar k\Delta_0 e^{-i\theta}\\
    v\hbar k\Delta_0 e^{i\theta}&0&(\beta-\alpha)/2&v\hbar k e^{-i\theta}\\
    0&v\hbar k\Delta_0 e^{i\theta}&v\hbar k e^{i\theta}&(\alpha+\beta)/2-\hbar\Omega
    \end{pmatrix}\begin{pmatrix}
    b_{1}(\boldsymbol{k})\\
    b_{2}(\boldsymbol{k})\\
    b_{3}(\boldsymbol{k})\\
    b_{4}(\boldsymbol{k})
    \end{pmatrix}.
\end{equation}
Considering a non-trivial solution in the previous equation, we arrive at the gapped quasienergy spectrum for Kek-Y distorted graphene under electromagnetic radiation
\begin{eqnarray}
 \varepsilon_{\mathrm{Y}}^{\xi,\eta}(k)=\eta\frac{1}{2}\Big(\sqrt{(\alpha-\hbar\Omega)^2+(2v\hbar k)^2}
 +\xi\sqrt{(\beta-\hbar\Omega)^2+(2\Delta_0v\hbar k)^2}\Big),
\end{eqnarray}
and the final four-components spinors of Eq. \eqref{ansatz} have the following form
\begin{equation}\label{spinors}
    \boldsymbol{\Psi}_{\xi,\eta}(\boldsymbol{k},t)=e^{-i\varepsilon_{\mathrm{Y}}^{\xi,\eta} t/\hbar}
    \begin{pmatrix}
    b_{1}^{\xi,\eta}(\boldsymbol{k})\\
    b_{2}^{\xi,\eta}(\boldsymbol{k})\\
    b_{3}^{\xi,\eta}(\boldsymbol{k})\\
    b_{4}^{\xi,\eta}(\boldsymbol{k})
    \end{pmatrix},
\end{equation}
where
\begin{eqnarray}
 b_{1}^{\xi,\eta}(\boldsymbol{k})&=&\mathcal{N}e^{-i2\theta}\Delta_0\left\{\big[\varepsilon_{\mathrm{Y},-}^g/2-\varepsilon_{\mathrm{Y}}^{\xi,\eta}(k)\big]\big[\varepsilon_{\mathrm{Y},+}^g/2-\varepsilon_{\mathrm{Y}}^{\xi,\eta}(k)\big]-(\Delta_0^2-1)(v\hbar k)^2\right\},\\
 b_{2}^{\xi,\eta}(\boldsymbol{k})&=&\mathcal{N}2e^{-i\theta}\Delta_0v\hbar k\varepsilon_{\mathrm{Y}}^{\xi,\eta}(k),\\
 b_{3}^{\xi,\eta}(\boldsymbol{k})&=&\mathcal{N}\frac{e^{-i\theta}v\hbar k}{\varepsilon_{\mathrm{Y},-}^g/2+\varepsilon_{\mathrm{Y}}^{\xi,\eta}(k)}\left\{\big[\varepsilon_{\mathrm{Y},-}^g/2-\varepsilon_{\mathrm{Y}}^{\xi,\eta}(k)\big]\big[(\Delta_0^2-1)\varepsilon_{\mathrm{Y},+}^g/2-(\Delta_0^2+1)\varepsilon_{\mathrm{Y}}^{\xi,\eta}(k)\big]-(\Delta_0^2-1)^2(v\hbar k)^2\right\},\\
 b_{4}^{\xi,\eta}(\boldsymbol{k})&=&\mathcal{N}\left\{-\big[\varepsilon_{\mathrm{Y},-}^g/2-\varepsilon_{\mathrm{Y}}^{\xi,\eta}(k)\big]\big[\varepsilon_{\mathrm{Y},+}^g/2+\varepsilon_{\mathrm{Y}}^{\xi,\eta}(k)\big]+(\Delta_0^2-1)(v\hbar k)^2\right\},
\end{eqnarray}
and $\mathcal{N}$ a normalization constant. It is important to remark here that in absence of the external electromagnetic field, the spinors of Eq. \eqref{spinors} coincide with the solution of unperturbed Kek-Y distorted graphene \eqref{baseSatateSpinors}.

A similar procedure can be followed to compute the quasienergy spectrum for the $\nu=0$ Kek-O distorted graphene under circularly polarized light. The expression for this spectrum is given in Eq. \eqref{QuasiOCirc}.

\section{Linearly polarized light}\label{Append-B}

Similarly to the Appendix \ref{Append-A}, in this part we show the main results to obtain the quasienergies spectrum for the $\nu=1$ Kek-Y distorted graphene (with a real $\tilde{\Delta}=\Delta_0$) under linearly polarized light. From Eq. \eqref{NonStatDiracE} and using the vector potential \eqref{vectPotLin}, we find
\begin{eqnarray}
    i\hbar\frac{d}{dt}\boldsymbol{\psi}(t)&=&-\frac{eE_0v}{\Omega}\cos(\Omega t)\begin{pmatrix}
    0&1&\Delta_0 &0\\
    1&0&0&\Delta_0 \\
    \Delta_0 &0&0&1 \\
    0&\Delta_0 &1&0\\
    \end{pmatrix}
    \boldsymbol{\psi}(t).
\end{eqnarray}
The last expression can be easily solved by integration, hence the solution is
\begin{equation}\label{SpinorYlp}
\boldsymbol{\psi}_{n}(t)=\boldsymbol{\psi}_{\xi,\eta}(t)=
\frac{1}{2}e^{i\eta\tilde{E}(1+\xi\Delta_0)\sin(\Omega t)}
\begin{pmatrix}
\xi 1\\
\xi \eta 1\\
\eta 1\\
1
\end{pmatrix},
\end{equation}
with $\xi=\pm$, $\eta=\pm$, and $n\in\{1,2,3,4\}$ as in Eq. \eqref{ketsKekY}.

Now according to the Floquet theorem, Eq.~\eqref{ansatz} is a periodic function of time with period $T=2\pi/\Omega$. Therefore, we can express each coefficient $a_{j}(\boldsymbol{k},t)$ as a Fourier series
\begin{equation}\label{ajFourier}
   a_{j}(\boldsymbol{k},t)=\sum_{m=-\infty}^{\infty}e^{im\Omega t}f_{j,m}(\boldsymbol{k}), \qquad j\in\{1,2,3,4\}.
\end{equation}
with these coefficients, we can write the sum in Eq.~\eqref{ansatz} to later substitute into the Dirac equation Eq.~\eqref{DiracEquation} and obtain 
\begin{equation}\label{algebraicEq}
    \sum_{m=-\infty}^{\infty}\big[\varepsilon-m\hbar\Omega \big]e^{im\Omega t}f_{j,m}(\boldsymbol{k})=\sum_{n=1}^4\sum_{m=-\infty}^{\infty}e^{im\Omega t}f_{n,m}(\boldsymbol{k})\boldsymbol{\psi}_j^{\dagger}(t)H_0\boldsymbol{\psi}_n(t).
\end{equation}
To obtain the eigenvalues $\varepsilon$ from this expression, we expand the exponential in $\boldsymbol{\psi}_j$ and $\boldsymbol{\psi}_n$ (see Eq.~\eqref{SpinorYlp}), using the Jacobi-Anger expansion
\begin{equation}
e^{iz\sin(\phi)}=\sum_{s=-\infty}^{\infty}J_s(z)e^{is\phi},
\end{equation}
where $J_s(z)$ is the $s$-esim Bessel function of the first kind. After using the orthogonality condition in the time dependent exponentials, we get
\begin{equation}\label{system}
\big[\varepsilon-m\hbar\Omega\big]f_{j,m}(\boldsymbol{k})=\sum_{n=1}^4\left[\sum_{s=-\infty}^{\infty}\big(\mathcal{J}_s\big)_{jn}f_{n,m-s}(\boldsymbol{k})+\mathcal{K}_{jn}f_{n,m}(\boldsymbol{k})\right],
\end{equation}
where $\mathcal{K}_{jn}$ and  $\big(\mathcal{J}_s\big)_{jn}$ are the elements of the matrices
\begin{equation}
    \mathcal{K}=v\hbar k\cos(\theta)\begin{pmatrix}
    1+\Delta_0&0&0&0\\
    0&\Delta_0-1&0&0\\
    0&0&1-\Delta_0&0\\
    0&0&0&-(1+\Delta_0)
    \end{pmatrix},
\end{equation}
and 
\begin{equation}
  \mathcal{J}_s=iv\hbar k\sin(\theta)\begin{pmatrix}
  0&-J_s\big(-2\tilde{E}\big)&-\Delta_0J_s\big(-2\Delta_0\tilde{E}\big)&0\\
  J_s\big(2\tilde{E}\big)&0&0&-\Delta_0J_s\big(-2\Delta_0\tilde{E}\big)\\
  \Delta_0J_s\big(2\Delta_0\tilde{E}\big)&0&0&-J_s\big(-2\tilde{E}\big)\\
   0&\Delta_0J_s\big(2\Delta_0\tilde{E}\big)& J_s\big(2\tilde{E}\big)&0
  \end{pmatrix},
\end{equation}
respectively.

Here we focus on the high-frequency field, this condition implies $(\varepsilon/\hbar\Omega)\approx 0$ and $ (vk/\Omega)\approx 0$.
Then, Eq. \eqref{system} can be reduced as
\begin{equation}
    -m\hbar\Omega f_{j,m}(\boldsymbol{k})=\sum_{n=1}^4\sum_{s=-\infty}^{\infty}\big(\mathcal{J}_s\big)_{jn}f_{n,m-s}(\boldsymbol{k}).
\end{equation}

The normalization condition implies that $\left|f_{j,m}(\boldsymbol{k})\right|\leq1$, and the Bessel functions obey $|J_s(z)|\leq1$. Keeping this in mind, as previous reported\cite{kibis}, the last equation leads to $f_{j,m\neq0}\approx0$. Therefore
it is possible to neglect all the Fourier coefficients
except for $m=0$. Under these considerations, Eq. \eqref{system} is reduced to the following eigenvalue problem
\begin{equation}\label{eigenprobKekO}
    \Big(\varepsilon\:\mathbb{I}_{4\times4}-\mathcal{K}-\mathcal{J}_0\Big)\begin{pmatrix}
    f_{1,0}(\boldsymbol{k})\\
    f_{2,0}(\boldsymbol{k})\\
    f_{3,0}(\boldsymbol{k})\\
    f_{4,0}(\boldsymbol{k})
    \end{pmatrix}=0,
\end{equation}
where $\mathbb{I}_{4\times4}$ is the identity matrix. 

Finally, considering a non-trivial solution in Eq. \eqref{eigenprobKekO}, we arrive at the gappless quasienergy spectrum for the Kek-Y distorted graphene under linearly polarized light,
\begin{eqnarray}
    \varepsilon_{\mathrm{Y}}^{\xi,\eta}(\boldsymbol{k})=\eta v\hbar k\left(\sqrt{\cos^2{(\theta)}+J_0^2(2\tilde{E})\sin^2{(\theta)}}+\xi \Delta_0\sqrt{\cos^2{(\theta)}+J_0^2(2\Delta_0\tilde{E})\sin^2{(\theta)}}\right).
\end{eqnarray}


{

A similar procedure can be followed to compute the quasienergy spectrum for the $\nu=0$ Kek-O distorted graphene under linearly polarized light. The expression for this spectrum is shown in  Eq. \eqref{QuasiOLin}.
}

\section{dc conductivity}\label{Append-C}

{
In this Appendix, we show the relevant calculations to obtain the current density, the relaxation time, and the conductivity for the Kek-Y distorted graphene under circularly polarized light.

First, in order to calculate the dc conductivity we introduce a stationary electric field to the Kekul\'e-distorted graphene sample. For simplicity, we assume an electric field along the $\hat{\boldsymbol{x}}$ direction $\boldsymbol{E}=(E_x,\:0)$. The Boltzmann equation for current density at zero temperature and a Fermi energy $\varepsilon_F$ in the conduction band ($\eta=+$) is given by
\begin{equation}\label{current}
    \boldsymbol{J}=\sum_{\xi}\frac{e^2}{(2\pi)^2}\int d^2\boldsymbol{k}\:\big(\boldsymbol{E}\cdot\boldsymbol{v}_{\xi}(k)\big)\tau_{\xi}(\boldsymbol{k})\boldsymbol{v}_{\mathrm{}}(k)\delta\big(\varepsilon_{\mathrm{Y}}^{\xi,+}(k)-\varepsilon_F\big),
\end{equation}
where $\varepsilon_F$ is the Fermi energy, $\boldsymbol{v}_{\xi}(k)=(1/\hbar)\nabla_{\boldsymbol{k}}\varepsilon_{\mathrm{Y}}^{\xi,+}(k)$ is the group velocity, and $\tau_{\xi}(\boldsymbol{k})$ is the relaxation time, which in the more general case in polar coordinates is given by\cite{relaxTime}
\begin{equation}\label{relax}
    \frac{1}{\tau_{\xi}(\boldsymbol{k})}=\sum_{\xi'}\frac{A}{(2\pi)^2}\int d\theta'\int k'dk'\left[1-\frac{\tau_{\xi'}({\boldsymbol{k'})}\big(\boldsymbol{E}\cdot\boldsymbol{v}_{\xi'}(k')\big)}{\tau_{\xi}(\boldsymbol{k})\big(\boldsymbol{E}\cdot\boldsymbol{v}_{\xi}(k)\big)}\right]w_{\boldsymbol{k}'\boldsymbol{k}}^{\xi',\xi},
\end{equation}
where $w_{\boldsymbol{k}'\boldsymbol{k}}^{\xi',\xi}$ is the probability of horizontal transitions of conduction electrons between the cone $\xi$ (initial state) with wave vector $\boldsymbol{k}$, and the cone $\xi'$ (final state) with wave vector $\boldsymbol{k}'$, per unit time. The sum in Eq. \eqref{current} runs over the initial states, and the sum in Eq. \eqref{relax} runs over the final states.

Second, we calculate $w_{\boldsymbol{k}'\boldsymbol{k}}^{\xi',\xi}$  using the four-component spinors of the Kek-Y distorted graphene under circularly polarized light \eqref{spinors}}
\begin{equation}\label{prob}
    w_{\boldsymbol{k}'\boldsymbol{k}}^{\xi',\xi}=\frac{2\pi}{\hbar}|\chi_{\boldsymbol{k}'\boldsymbol{k}}^{\xi',\xi}|^2|U_{\boldsymbol{k}'\boldsymbol{k}}|^2\delta\big(\varepsilon_{\mathrm{Y}}^{\xi',+}(k')-\varepsilon_{\mathrm{Y}}^{\xi,+}(k)\big),
\end{equation}
where 
\begin{equation}\label{braket}
    \chi_{\boldsymbol{k}'\boldsymbol{k}}^{\xi',\xi}=\sum_{n=1}^{4}b_{n}^{\xi,+}(\boldsymbol{k})\big(b_{n}^{\xi',+}(\boldsymbol{k}')\big)^{\ast},
\end{equation}
such that
\begin{eqnarray}\label{xi}
   \nonumber|\chi_{\boldsymbol{k}'\boldsymbol{k}}^{\xi',\xi}|^2&=&\Big[|b_{2}^{\xi',+}(k')||b_{2}^{\xi,+}(k)|+|b_{3}^{\xi',+}(k')||b_{3}^{\xi,+}(k)|\Big]^2+|b_{1}^{\xi',+}(k')|^2|b_{1}^{\xi,+}(k)|^2+|b_{4}^{\xi',+}(k')|^2|b_{4}^{\xi,+}(k)|^2\\ \nonumber&&+2\Big[|b_{2}^{\xi',+}(k')||b_{2}^{\xi,+}(k)|+|b_{3}^{\xi',+}(k')||b_{3}^{\xi,+}(k)|\Big]|b_{1}^{\xi',+}(k')||b_{1}^{\xi,+}(k)|\cos(\theta'-\theta)\\
   \nonumber&&+2|b_{4}^{\xi',+}(k')||b_{4}^{\xi,+}(k)|\Big\{\Big[|b_{2}^{\xi',+}(k')||b_{2}^{\xi,+}(k)|+|b_{3}^{\xi',+}(k')||b_{3}^{\xi,+}(k)|\Big]\cos(\theta'-\theta)
   +|b_{1}^{\xi',+}(k')||b_{1}^{\xi,+}(k)|\cos(2\theta'-2\theta)\Big\},\\
\end{eqnarray}
and $|U_{\boldsymbol{k}'\boldsymbol{k}}|^2=\frac{N_{A}}{A}U_0^2$ according to Eq. \eqref{ScatPot}. Now, considering that the relaxation time does not depend upon the angle $\theta$, from Eq. $\eqref{relax}$ we can find two relaxation times. If we consider the internal cone as the initial state ($\xi=+$) we obtain
\begin{eqnarray}\label{algEq1}
   \frac{1}{\tau_+(k)}&=&\frac{N_AU_0^2}{\hbar^2}\frac{k^-}{v_-(k^-)}\left[\Gamma_1(k,k^-)-\frac{\tau_-(k^-)v_-(k^-)}{\tau_+(k)v_+(k)}\Gamma_2(k,k^-)\right]+\frac{N_AU_0^2}{\hbar^2}\frac{k}{v_+(k)}\Gamma^{+}_3(k),
\end{eqnarray}
and if now we chose the external cone as the initial state ($\xi=-$), 
\begin{eqnarray}\label{algEq2}
   \frac{1}{\tau_-(k)}&=&\frac{N_AU_0^2}{\hbar^2}\frac{k^+}{v_+(k^+)}\left[\Gamma_1(k^+,k)-\frac{\tau_+(k^+)v_+(k^+)}{\tau_-(k)v_-(k)}\Gamma_2(k^+,k)\right]+\frac{N_AU_0^2}{\hbar^2}\frac{k}{v_-(k)}\Gamma^{-}_3(k),
\end{eqnarray}
where
\begin{eqnarray}
\Gamma_1(\kappa,\tilde{\kappa})&=&\Big(|b_{2}^{-,+}(\tilde{\kappa})||b_{2}^{+,+}(\kappa)|+|b_{3}^{-,+}(\tilde{\kappa})||b_{3}^{+,+}(\kappa)|\Big)^2+|b_{1}^{-,+}(\tilde{\kappa})|^2|b_{1}^{+,+}(\kappa)|^2+|b_{4}^{-,+}(\tilde{\kappa})|^2|b_{4}^{+,+}(\kappa)|^2,\\
\Gamma_2(\kappa,\tilde{\kappa})&=&\Big(|b_{2}^{+,+}(\kappa)||b_{2}^{-,+}(\tilde{\kappa})|+|b_{3}^{+,+}(\kappa)||b_{3}^{-,+}(\tilde{\kappa})|\Big)\Big(|b_{1}^{+,+}(\kappa)||b_{1}^{-,+}(\tilde{\kappa})|+|b_{4}^{+,+}(\kappa)||b_{4}^{-,+}(\tilde{\kappa})|\Big),\\
\nonumber\Gamma_3^{\pm}(\kappa)&=&\Big(|b_{2}^{\pm,+}(\kappa)|^2+|b_{3}^{\pm,+}(\kappa)|^2\Big)^2+|b_{1}^{\pm,+}(\kappa)|^4+|b_{4}^{\pm,+}(\kappa)|^4-\Big(|b_{2}^{\pm,+}(\kappa)|^2+|b_{3}^{\pm,+}(\kappa)|^2\Big)\Big(|b_{1}^{\pm,+}(\kappa)|^2+|b_{4}^{\pm,+}(\kappa)|^2\Big).\\
\end{eqnarray}
We defined the wave vectors $k^+$ such that $\varepsilon_{\mathrm{Y}}^{+,+}(k^+)=\varepsilon_{\mathrm{Y}}^{-,+}(k)$, and  $k^-$ with $\varepsilon_{\mathrm{Y}}^{-,+}(k^-)=\varepsilon_{\mathrm{Y}}^{+,+}(k)$. The explicit forms of $k^{+}$ and $k^{-}$ are given by the following expression
\begin{eqnarray}
    \nonumber k^{\pm}=\frac{1}{v\hbar(1-\Delta_0^2)}\Big\{\mp\varepsilon_{\mathrm{Y}}^{\mp,+}(k)\sqrt{4\Delta_0^2\big(\varepsilon_{\mathrm{Y}}^{\mp,+}(k)\big)^2+(1-\Delta_0^2)\big[(\beta-\hbar\Omega)^2-(\alpha-\hbar\Omega)^2\Delta_0^2\big]}\\+\:(1+\Delta_0^2)\big(\varepsilon_{\mathrm{Y}}^{\mp,+}(k)\big)^2-(\varepsilon_{\mathrm{Y},-}^g/2)(\varepsilon_{\mathrm{Y},+}^g/2)(1-\Delta_0^2)\Big\}^{1/2},\qquad\qquad
\end{eqnarray}
For a fixed Fermi energy $\varepsilon_F$, the Fermi wave vectors $k_F^+$ and $k_F^-$ are defined such that $\varepsilon_{\mathrm{Y}}^{+,+}(k_F^+)=\varepsilon_{\mathrm{Y}}^{-,+}(k_F^-)=\varepsilon_F$. Then, we notice that Eqs. \eqref{algEq1} and \eqref{algEq2} form a system of algebraic equations for $\tau_{+}(k_F^+)$ and  $\tau_{-}(k_F^-)$, that is to say
\begin{eqnarray}\label{algEq}
   \nonumber\frac{1}{\tau_+(k_F^+)}&=&\frac{N_AU_0^2}{\hbar^2}\frac{k^-}{v_-(k_F^-)}\left[\Gamma_1(k_F^+,k_F^-)-\frac{\tau_-(k_F^-)v_-(k_F^-)}{\tau_+(k_F^+)v_+(k_F^+)}\Gamma_2(k_F^+,k_F^-)\right]+\frac{N_AU_0^2}{\hbar^2}\frac{k_F^+}{v_+(k_F^+)}\Gamma_3(k_F^+),\\
   \frac{1}{\tau_-(k_F^-)}&=&\frac{N_AU_0^2}{\hbar^2}\frac{k_F^+}{v_+(k_F^+)}\left[\Gamma_1(k_F^+,k_F^-)-\frac{\tau_+(k_F^+)v_+(k_F^+)}{\tau_-(k_F^-)v_-(k_F^-)}\Gamma_2(k_F^+,k_F^-)\right]+\frac{N_AU_0^2}{\hbar^2}\frac{k_F^-}{v_-(k_F^-)}\Gamma_3(k_F^-).
\end{eqnarray}
By solving the system we obtain,
\begin{eqnarray}\label{tauPlus}
    \nonumber\tau_{+}(k_F^+)&=&\frac{\hbar^2}{N_AU_0^2}v_{-}(k_F^{-})v_{+}(k_F^+)\Big\{v_{-}(k_F^{-})\big[k_F^+\Gamma_1(k_F^+,k_F^{-})+k_F^{-}\Gamma_2(k_F^+,k_F^-)\big]+v_{+}(k_F^+)k_F^{-}\Gamma_3^-(k_F^-)\Big\}\\\nonumber&&\times\Big\{\big[v_{+}(k_F^+)k_F^{-}\Gamma_1(k_F^+,k_F^-)+v_{-}(k_F^{-})k_F^+\Gamma_3^+(k_F^+)\big]\big[v_{-}(k_F^{-})k_F^+\Gamma_1(k_F^+,k_F^-)+v_{+}(k_F^+)k_F^{-}\Gamma_3^-(k_F^-)\big]\\&&\quad+\:v_{+}(k_F^+)v_{-}(k_F^{-})k_F^+k_F^{-}\big[\Gamma_2(k_F^+,k_F^-)\big]^2\Big\}^{-1},
\end{eqnarray}
and
\begin{eqnarray}\label{tauMinus}
    \nonumber\tau_{-}(k_F^-)&=&\frac{\hbar^2}{N_AU_0^2}v_{-}(k_F^-)v_{+}(k_F^{+})\Big\{v_{+}(k_F^{+})\big[k_F^-\Gamma_1(k_F^{+},k_F^-)+k_F^{+}\Gamma_2(k_F^{+},k_F^-)\big]+v_{-}(k_F^-)k_F^+\Gamma_3^+(k_F^{+})\Big\}\\\nonumber&&\times\Big\{\big[v_{-}(k_F^-)k_F^+\Gamma_1(k_F^{+},k_F^-)+v_{+}(k_F^+)k_F^-\Gamma_3^-(k_F^-)\big]\big[v_{+}(k_F^{+})k_F^-\Gamma_1(k_F^{+},k_F^-)+v_{-}(k_F^-)k_F^{+}\Gamma_3^+(k_F^{+})\big]\\&&\quad+\:v_{-}(k_F^-)v_{+}(k_F^{+})k_F^-k_F^{+}\big[\Gamma_2(k_F^{+},k_F^-)\big]^2\Big\}^{-1}.
\end{eqnarray}
Now, we consider the $J_x$ component of the current density in polar coordinates
\begin{equation}
    J_x=\sum_{\xi}\frac{e^2E_x}{(2\pi)^2}\int_0^{2\pi} d\theta\int_0^{\infty} kdk\:\tau_{\xi}(k)v_{\xi}^2(k)\cos^2(\theta)\delta\big(\varepsilon_{\mathrm{Y}}^{\xi,\eta}(k)-\varepsilon_F\big).
\end{equation}
By solving the last integral we obtain
\begin{equation}
    J_x=\sum_{\xi}\frac{e^2E_x}{2\pi\hbar} k_F^{\xi}\tau_{\xi}(k_F^{\xi})v_{\xi}(k_F^{\xi}),
\end{equation}
where we have considered the spin degeneracy (factor 2). Finally the dc conductivity is given by $\sigma=J_x/E_x$ (see Eq. \eqref{finalsigma}).

\twocolumngrid

\bibliography{referencias.bib}

\end{document}